# Clafer: Lightweight Modeling of Structure, Behaviour, and Variability

Paulius Juodisius[a], Atrisha Sarkar[b], Raghava Rao Mukkamala[c], Michał Antkiewicz[b], Krzysztof Czarnecki[b], and Andrzej Wąsowski[a]

- a    IT University of Copenhagen, Denmark
- b    GSD Lab, University of Waterloo, Canada
- c    Department of Technology, Kristiania University College, Norway

**Abstract**    Embedded software is growing fast in size and complexity, leading to intimate mixture of complex architectures and complex control. Consequently, software specification requires modeling both structures and behaviour of systems. Unfortunately, existing languages do not integrate these aspects well, usually prioritizing one of them. It is common to develop a separate language for each of these facets.

In this paper, we contribute Clafer: a small language that attempts to tackle this challenge. It combines rich structural modeling with state of the art behavioural formalisms. We are not aware of any other modeling language that seamlessly combines these facets common to system and software modeling.

We show how Clafer, in a single unified syntax and semantics, allows capturing feature models (variability), component models, discrete control models (automata) and variability encompassing all these aspects. The language is built on top of first order logic with quantifiers over basic entities (for modeling structures) combined with linear temporal logic (for modeling behaviour). On top of this semantic foundation we build a simple but expressive syntax, enriched with carefully selected syntactic expansions that cover hierarchical modeling, associations, automata, scenarios, and Dwyer's property patterns.

We evaluate Clafer using a power window case study, and comparing it against other notations that substantially overlap with its scope (SysML, AADL, Temporal OCL and Live Sequence Charts), discussing benefits and perils of using a single notation for the purpose.

**ACM CCS 2012**

- **Software and its engineering** → **Formal language definitions**; **System description languages**; **Design languages**; **Specification languages**;

## The Art, Science, and Engineering of Programming

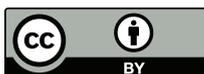

**Clafer: Lightweight Modeling of Structure, Behaviour, and Variability**

# 1 Introduction

A multitude of languages and tools supports either modeling structure, or behavior, or variability. These three aspects seem to be key to software and system modeling. We have structural modeling languages (say UML class diagrams), architectural modeling languages (SysML, AADL, EAST-ADL), behavioural languages (statecharts, UML state diagrams, Simulink stateflow), and variability languages (feature models, OVMs or CVL). The ability to combine these aspects is crucial, as complexity of software increases, bringing rich architecture to almost any system. Many of the above languages do include facilities for combining structure, variability, and behaviour, but typically using separate notations. For instance, all architecture description languages support variability modeling, and the UML combines a variety of diagrams. OVM and CVL are mechanisms to introduce variability to any structural (or behavioural) language. Our experience, however, observing how these combinations are used in teaching, engineering, and research, as well as the experience from designing some of them, indicates that multiple notations are cognitively hard to work with; See for instance, by now, cliché problems in UML caused by complexity [45, 46]; similar problems can be found in many works about semantics of the UML. Models combining multiple notations are difficult to comprehend holistically, both because of limitations of human cognition and errors in language design that accumulates with complexity.

Furthermore, the separation of notations leads to compartmentalization of the modeling process, which, while often being useful, can also be distracting. For example, it is very difficult to think of variability modeling as a separate concern from structural modeling—after all, variability in structure is a generalization of structural modeling, and not an entirely separate activity. However in reality, the information about structure, behaviour, and variability becomes available in parallel during the requirements elicitation process. A language where variability is integrated into structural modeling (or behavioral modeling) allows the user to choose whether to build two separate models or not. A language, where these aspects are separated at language design stage, does not offer such choice.

For these reasons, it is interesting to experiment with designing new integrated languages that would lower barrier of entry and counteract (the unproductive) compartmentalization of modeling, while allowing it when needed. Over the years, we have collected the experience of using and studying various modeling languages in the space of system modeling [8, 9, 10, 11, 13, 19, 23, 32, 39, 53, 59, 71]. We have learnt that minimality of concepts helps popularity of simple modeling styles like feature models [10, 19, 39]; and that more complex variability spaces require modeling architectural structures [8, 11], including channels, ports, and components [32]. We have seen uses of both direct behavioral modeling (automata style) [23, 71], and the advantages of declarative modeling of behavior [53]. Finally, we knew that the structure and behavior need to be mappable to high level variability descriptions like feature models and decision models [9, 39, 59]. We used this accumulated experience to design a language that combines modeling of both structure and behaviour with variability using a small number of concepts. The language follows declarative semantics, but embeds syntax that allows models to appear similarly to automata (state-based).

Settling on very few first-class concepts shared by multiple modeling styles is key to avoid building a union of languages like the UML, while still maintaining the ability to cover many applications.

Not only our own, but the accumulated experience of the entire research community with a multitude of languages gives an excellent empirical basis for understanding which concepts and mechanisms are key to modeling software systems. Many concepts are shared across behavioural and structural languages, including kind-of/part-of decomposition (seen in feature models, class diagrams and all architectural languages), inheritance/specialization/extension (in state diagrams, class diagrams and in architectural languages), variation points (in variability languages and architectural languages), or hierarchical nesting (almost everywhere). Most of the languages incorporate *looseness* mechanisms—an under-constraining used to express variability, dynamicity, abstraction, approximation or uncertainty.

Our hypothesis is that a language combining entities and relationships, refinement, hierarchical nesting, and looseness will be very compact while still capable of expressing multiple aspects in a single notation. We believe that such a language, while not universal, would be reasonably comprehensive and easier to learn and use than more specialized languages that over time accumulated accidental complexity growing into large, complex and heterogeneous notations.

We approach this objective by extending Clafer [4], an existing minimalistic language for structural modeling, with a temporal dimension; we simply interpret a structural Clafer model as having unrestricted behaviour: instances become snapshots, constraints become invariants, and all possible evolutions of instances are allowed. We extend the constraint language with linear temporal logics [57] to constrain the possible evolutions of instances in traces. We enrich the notation with syntactic sugar for Dwyer's property patterns [25], and for transitions for modeling hierarchical state machines [2, 36] and scenarios. Crucially, we do not add any new first-class elements to the language, which already supports basic entities (clafers),[1] part-of decomposition, inheritance, references, and constraints. The only extension are the temporal operators added to the constraint language and the trace-based interpretation in the semantics. To address the multitude of applications, we propose a collection of modeling patterns relying on existing modeling constructs. In short, the paper offers the following contributions:

- *The design of Clafer, a modeling language that combines behaviour, structure, and variability, including a formal specification of the core language.* To our best knowledge no single language exists that addresses this requirement.
- *A demonstration of modeling patterns for structure, behaviour, and variability.* As far as we know, this is the first, reasonably systematic, approach to collect such patterns, instead of demanding language extensions to achieve similar functionality.

---

[1] We capitalize *Clafer* when referring to the language, and use small caps to refer to *clafers*, its basic entities, which roughly correspond to classes, attributes and references. From now on we write "Clafer" meaning "Clafer extended with behaviour".

- *An implementation of the language in the Clafer compiler* capable of desugaring the extended syntactic elements (transitions and property patterns) into the basic temporal logic expressions.[2] The parser is generated from a formal grammar specification, whereas the desugarer is implemented manually.
- *A handful of publicly available models*; few models combining variability and behaviour have been freely available before, despite their growing importance.

Clafer with behaviour combines an established structural modeling paradigm (class diagrams) with a solid semantic model for behaviour (traces). It allows creation of "loose" models (underspecified, with variability, with uncertainty), and mixing looseness of the different aspects. It does not require compartmentalization of behaviour and structure, but allows it. It allows organizing behavioural specifications via structural modeling concepts. Finally, it allows introducing variability uniformly in all the above.

We believe that a small language capturing diverse viewpoints can facilitate model driven development of high quality software in more domains and at lower cost than today. We also hope that it will influence design of future industry standards. A small language with unambiguous semantics makes it easier to develop formal analysis tools for models mixing structure, behaviour and variability. Last but not least, we believe that Clafer has good potential for teaching modeling. A small language can be learned faster. Students can experience the different facets of modeling quicker, and obtain operational skills in different kinds of modeling using a single language, which is important in a short time horizon of a course. It can demonstrate a variety of patterns and use cases without bringing the ballast of multiple notations and tools into the learning activities.

We proceed by introducing the running example in section 2. We show patterns for feature modeling, structural modeling, and variability modeling in section 3, followed by modeling behaviour including control, state machines, properties, scenarios and variability in section 4. We present the semantics of the core Clafer in section 5 and discuss additional constructs and the implementation in section 6. Similarly, section 7 discusses the design and properties of Clafer against the baseline of several other languages, in an attempt to evaluate this work using a case study. We gather the lessons learnt in section 8, reflecting on the design and giving pointers to future work. We discuss related work (section 9) and conclude in section 10.

## 2  The Running Example

We use a power window subsystem of a contemporary car as a running example, a choice inspired by a realistic case study, which we pursued in collaboration with an industrial partner in the automotive domain [54, 62, 63].[3] An electrically-powered window receives inputs from the users that request moving the window glass up (to

---

[2] Available from http://www.clafer.org, last accessed on 2018-07-19.
[3] https://github.com/gsdlab/ClaferCaseStudies/, last accessed on 2018-07-19.

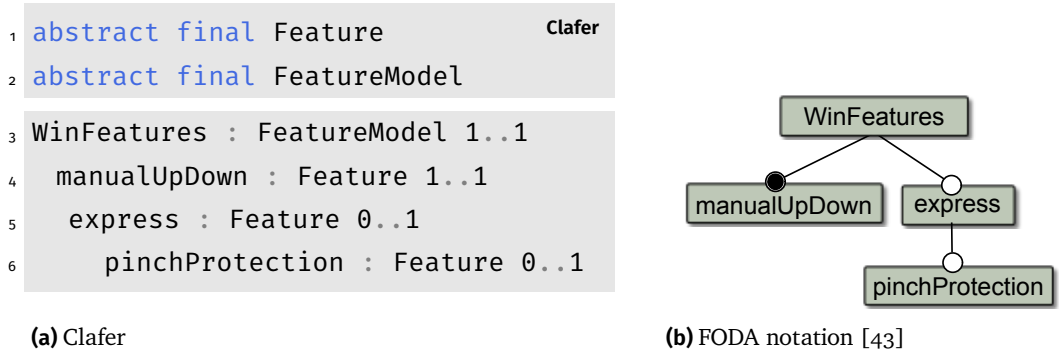

```
1  abstract final Feature                    Clafer
2  abstract final FeatureModel
3  WinFeatures : FeatureModel 1..1
4    manualUpDown : Feature 1..1
5     express : Feature 0..1
6       pinchProtection : Feature 0..1
```

**(a)** Clafer    **(b)** FODA notation [43]

■ **Figure 1** An example feature model. The Clafer model (left) defines both the types representing the core concepts of feature modeling and the feature model itself.

close) or to close it automatically (express up), to move it down (open), and to stop the movement of the glass panel. The window controller actuates the window motor to execute these commands. However, up commands are ignored when the window is closed, and down commands are ignored when the window is completely open, to avoid burning the motor. Finally, a safety feature (pinch protection) suspends the closing movement if an obstacle is detected.

The example is sufficiently rich to discuss several modeling viewpoints. It is also sufficiently complex to illustrate the need for a compact notation and a well-defined semantics supporting building automatic analysis tools, which could help modelers. To facilitate comparison, we include fragments of the example in dominant modeling languages for the different viewpoints. We start with demonstrating using Clafer for feature modeling and structural modeling. While, these applications of Clafer were previously presented by Bąk, Diskin, Antkiewicz, Czarnecki, and Wąsowski [4] in an entirely static setting; here we introduce explicit patterns for obtaining different kinds of models as well as we take a dynamic perspective, which allows some new aspects to appear (like the modifier `final`, or modeling of messages).

## 3   Background. Modeling Features and Structures with Variability

We begin the technical developments by introducing structural modeling capabilities of Clafer. While these aspects of the language have been known previously, we give an entirely new exposition to them, emphasizing the use of patterns in modeling. We will continue this style in section 4, where patterns for behavioral modeling will be introduced.

### 3.1 Feature Modeling

Feature models [43] are a notation for capturing key characteristics of a system from a high level requirements perspective. They are among the simplest and most abstract models created for software artifacts. They name the core common and variable features of a system, organize them into a hierarchy, and capture dependencies among them. A feature model thus describes the set of valid *feature configurations*—legal selections of features that can be combined to build a product variant. The set of all variants and a method to derive them is typically called a *product line*. Feature modeling has been popular for its simplicity and similar notations have been introduced in commercial[4] and open source tools [10, 56] aimed at building product lines and highly configurable systems.

Clafer (the language) has a minimalistic syntax that unifies *class*, *reference* and *property* into a single concept called *clafer* (type). From feature modeling point of view, a clafer also unifies both the *feature* and *feature group* concepts. In general, a model in Clafer is built from clafers that represent domain concepts and relations among them. Moreover, figure 1a presents a simple Clafer model. Each line contains one declaration of a clafer. The model can be interpreted as an equivalent feature model in the FODA notation (figure 1b).

Our family of Power Window systems only admits three variants. Every variant will contain the mandatory feature `manualUpDown` which corresponds to the basic functionality of a power window. Two optional features `express` and `pinchProtection` can also be selected for a variant configuration; however, `pinchProtection` can only be selected if the feature `express` is selected.

The Clafer model in figure 1a first declares the core types of feature modeling. The first two clafers are abstract, meaning that they cannot have direct instances of their own, but only via concrete clafers which extend them via generalization relation (akin to abstract classes in OO languages). The default multiplicity of abstract clafers is 0..* and we always omit it for clarity. In lines 3–5 the feature tree is specified using indentation (which means containment composition in Clafer). The root node represents the model, and the indented clafer declarations represent children. The intervals (`a..b`) following the declarations denote *clafer multiplicity* which constrain the cardinality of instance sets. Here the existence of an instance of the clafers `WinFeatures` and `manualUpDown` is mandatory (exactly one instance is required), while the two other features are optional (at most one instance is allowed). In FODA, optionality is represented using hollow dots and the root is mandatory by default. To summarize, the multiplicity of a clafer specifies how many instances of that clafer can be nested under an instance of its parent clafer. By using multiplicity 1..1 and 0..1, we can express mandatory and optional features, respectively. The presence of an instance of a clafer represents the fact that a feature is selected in a feature configuration. The nesting of clafers indicates that an instance of a child clafer can only exist nested under an instance of its parent clafer, which we use to express that a child feature can only be selected if its parent feature is also selected. In our example, we assume that configuration of the power window is performed statically (say at factory construction time). Thus, we would like to declare all clafers in the feature model as `final`—meaning that their instances cannot change (appear and disappear) once they are created. Instead of changing each clafer into final, we mark the abstract clafers `Feature` and `FeatureModel` as final so that this property is inherited by all clafers

---

[4] Available among others from Pure Systems GmbH and Big Lever Software Inc.

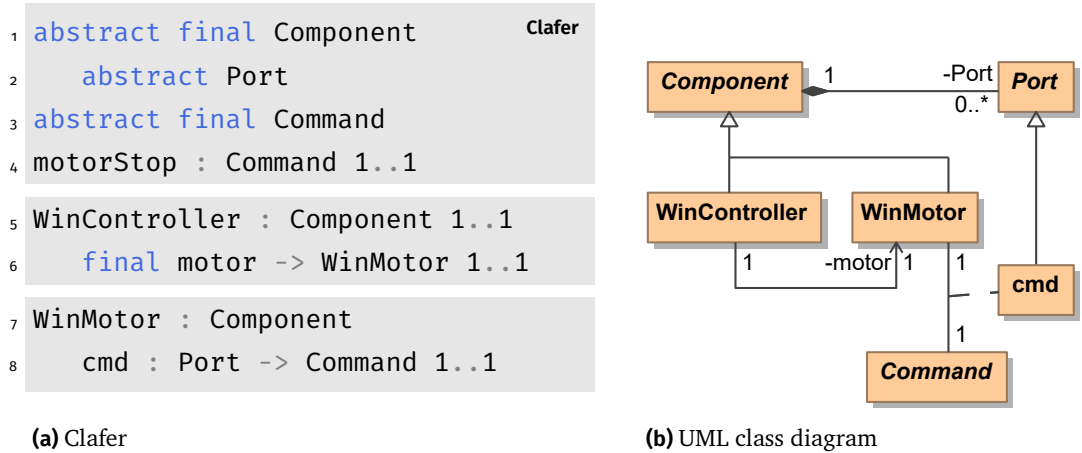

```
1  abstract final Component                    Clafer
2      abstract Port
3  abstract final Command
4  motorStop : Command 1..1
5  WinController : Component 1..1
6      final motor -> WinMotor 1..1
7  WinMotor : Component
8      cmd : Port -> Command 1..1
```

**(a)** Clafer                    **(b)** UML class diagram extending them. This is not visible in FODA notation, as it is rarely used in dynamic interpretation.

**Figure 2** A structural model of the core architecture of the Power Window example

In general, Clafer subsumes cardinality-based feature models [20], which in turn subsume FODA. It allows modeling both optional and mandatory features, features that can have multiple instances, feature groups (including and, or, xor and mutex groups), feature references and cross-tree constraints (such as requires and excludes). These features are not used in our example for simplicity. Yet, we will use the corresponding Clafer constructs in later examples. Note that feature modeling in Clafer does not use any first class feature modeling constructs. Instead a modeling pattern is used, in which nesting is used to represent feature hierarchy and cardinality constraints are used to model optionality and feature groups. Constraints can be written to capture further relations among features.

### 3.2 Structural Modeling

Architectural models are concerned with functional decomposition of features into structures of components that will realize the intended functionality. The key aspects of architectural modeling include identifying components, ports, and their relationships such as connections, containment (part-of) and refinement (kind-of). Known architectural modeling languages include AADL,[5] (originally for avionics applications) and EAST-ADL[6] (automotive domain). Architectural modeling languages are often encoded as profiles on class diagrams.

We shall now name and relate the main components of our example (see figure 2a). We begin by declaring the abstract clafers representing key concepts of architectural modeling: components and ports (lines 1–2). The clafer Command will represent the type of messages sent between the controller and the motor (e.g., we include an example concrete command motorStop on line 4).

---

[5] http://www.aadl.info/, last accessed on 2018-07-19.
[6] http://www.east-adl.info/, last accessed on 2018-07-19.

**Clafer: Lightweight Modeling of Structure, Behaviour, and Variability**

Lines 5–6 introduce the clafers representing the controller. The clafer `WinController`, which is a component, contains a reference clafer (`motor`) to exactly one instance of the clafer `WinMotor`, representing the motor (lines 7–8). We model the connection between the two using the reference `WinMotor.cmd`: the signals from the controller to the motor will be passed by setting the reference (e.g., `WinMotor.cmd=motorStop`, since both are singleton sets, the only instance of `cmd` must point to the only instance of `motorStop`).

The declaration of `cmd` contains both a clafer type (`: Port`) and a referenced type (`-> Command`). Both are optional in Clafer, as seen above (cf. `motorStop` and `motor`). The clafer type annotation indicates that a clafer specializes another clafer, analogously to class generalization in object-oriented modeling (for instance if a super-clafer contains properties, these are automatically included in the declared clafer). The reference type annotation restricts the type of referenced objects. Each clafer can be both a typed property (in the sense of a class) and a reference. The reference is "built-in" which allows using more lightweight syntax than in UML, when additional model element needs to be introduced and named to use references. If the reference type is omitted, then the built-in reference of the declared clafer is always empty. In this particular example, the use of `Port` has only a minor semantic impact: We use generalization annotation to clarify the role of the `cmd` clafer in the model (that it is a port). No properties are inherited, because `Port` is an empty type.

Observe that all `Components` and `motor` are final clafers (the architecture is static in our example), but `cmd` is not, which means that it will change over time.

A class diagram capturing the same information is shown in figure 2b. Clafer nesting is simply class containment (e.g., the class `Component` contains the class `Port`). Also, all clafers can have children, including the reference clafers. The two references, `motor` and `cmd`, otherwise similar in Clafer, have been modeled differently in the UML diagram. The former is captured as a simple reference, which cannot have attributes. The latter is *reified* (materialized) into an association class, which can have attributes and contain other classes. In fact, in Clafer, all references are always reified, so the second modeling (using the association class `cmd`) is more accurate. There are no special reference/association clafers in Clafer. Each reference is a clafer and each clafer can act as a reference. As all clafers can nest other clafers (representing properties), all clafer references can have attributes.

In general, the structural language of Clafer is very similar to MOF class diagrams. It supports inheritance (`:`) and references (`->` for set-valued references and `->>` for multiset-valued references), although the references are always reified and they can contain instances of other clafers. Each clafer can both point to a set of clafers and contain other clafers. Attributes are realized in the same simple framework, by referencing values of designated clafers representing simple types (thus, in a way, attributes are always boxed like in Smalltalk). For example, every instance of a clafer `length -> integer 1..1` must point to an instance of the clafer `integer`, which represents the set of all integer numbers.

Note that Clafer does not have any first class support for architectural concepts, but we follow a pattern when modeling an architecture. Like with Feature Models, we use abstract clafers to introduce components and ports. Rich connectors could have

■ **Listing 1** Relating a structural model and a feature model in Clafer

```Clafer
WinFeatures : FeatureModel 1..1
    manualUpDown : Feature 1..1
    express : Feature 0..1
        pinchProtection : Feature 0..1
WinController : Component 1..1
    final motor -> WinMotor 1..1
    pinchDetector : Component 0..1
    [ one this.pinchDetector <=>
        one WinFeatures.express.pinchProtection ]
```

been defined similarly. This is reminiscent of the SysML[7] practice of profiling UML for architectural use. Events and messages are modeled as clafers that are targets of a reference (or members of a set); the intuition being that of a typed slot that can be filled or not. As a result, the language with an extremely small core abstract syntax (see section 5) can express reasonably rich models.

### 3.3 Variability in Structural Models

Architectural modeling naturally requires variability modeling, whenever variation is involved, for instance when modeling a family of systems. One common way to add variability to architectural models is by linking the presence of architectural elements to conditions expressed in terms of features, so called *presence conditions*. We shall discuss it now, while alternatives will be shown in section 4. Presence conditions can be introduced directly in the modeling language [18], or by using an external notation, as in separate variability modeling, cf. Orthogonal Variability Modeling [58] (OVM) and the Common Variability Language[8] (CVL). In Clafer neither a separate notation, nor first-class presence conditions are needed.

In Clafer, we map elements to features using constraints. In listing 1, both the feature model and the component model are placed together (the abstract concept declarations and `WinMotor` are elided to conserve space). A new optional (0..1) component, `pinchDetector`, is introduced in line 7 as a part of the controller. In lines 8–10, a constraint states that an instance of the component `pinchDetector` is present if and only if an instance of the feature `pinchProtection` is present.

Clafer provides a constraint language that is largely inspired by Alloy's [42]. It is a language of relational set expressions combined with first-order quantifiers. Unlike in Alloy, where signatures cannot be nested, constraints in Clafer can be nested at any depth in a clafer hierarchy and under any clafer, including references. A constraint nested under a clafer is enforced for each instance of this clafer (so whenever such clafer exists). Technically speaking, the constraint is implicitly universally quantified over all instances of the context clafer. This roughly corresponds to use of contexts in the Object Constraint Language[9] (OCL). Ability to write constraints in different contexts, allows often writing them more compactly than at the top-level.

---

[7] http://sysml.org/, last accessed on 2018-07-19.
[8] http://variabilitymodeling.org/, last accessed on 2018-07-19.

When interpreting the model in a dynamic context, we would like all constraints in static Clafer to become invariants, that is, to *hold globally in a trace* as long as an instance of their context clafer exists. Although not shown in listing 1, every static constraint has an implicit temporal quantifier `globally` by default, which we explain in section 5.

Finally, observe that linking the structural model to the feature model has not relied on first-class notions of presence conditions, mappings, variation points, or traceability links (as in the above cited works). Instead a pattern was applied. We placed the feature model and a component model in a shared space, without any need for semantic integration, as they are written in the same language. For the same reason we could relate their elements using simple constraints, realizing the functionality of presence conditions. The obtained model is very concise. In contrast, applying an OVM or CVL style model requires creating three layers of artifacts (feature model, a mapping model of variation points and the base model—here the class diagram). Both require use of complex tools, as no easy syntax exists (CVL even lacks syntax for the mapping layer).

### 3.4 First-Order Constraints

In the early days of Clafer, we discussed the possibility of adaptation of constraint languages of various kinds, and performed semi-formal comparative studies of Alloy [42] and the Object-Constraint Language (OCL) [69]. Alloy's relational style was slightly more concise, and seemed to mix well with the keyword free-style of hierarchy modeling in Clafer.[10] The semantic adaptation was essentially direct, we just used clafers instead of Alloy's signatures as the main objects constrained. To make the paper self-contained, we summarize the key points of the resulting language below.

1. Constraints are written in square brackets.
2. A clafer name refers to the set of all instances appearing at a given point at runtime, being typed by this clafer. Thus `WinController` refers to all available window controllers. For clafers of cardinality one, like the window controller, this yields a singleton set. In general, larger sets, and empty sets, can arise for clafers with less constrained cardinality.
3. The semantics of the navigation dot is relational join with projection, like in Alloy. A top-level clafer `WinControllers` yields a set, while the constraint language interprets *nested* clafers as binary parent-child relations. Thus `pinchDetector` is seen as a binary relation between `WinControllers` and pinch detector instances.

---

[9] http://www.omg.org/spec/OCL/, last accessed on 2018-07-19.
[10] Jackson compares OCL with Alloy in the cited book and in the preceding papers.

    A navigation (`WinController.pinchDetector`) joins the relation with the set of `winControllers`, then drops the left column, leaving the set of pinch detectors that are nested in the given window controllers. While this semantics seems somewhat mathematical, it has been rather naturally adopted by Alloy users, since the syntax is intuitive—navigation dots follow semantics like in programming languages with structures or objects, just with the exception that one can navigate over more than a single instance at a time. In our experience, students pick up the expression language seamlessly, relying on experience from programming languages.

4. We allow dropping navigation and using just a single clafer name, if the name is unique in the model. This yields all instances of the given clafer, for instance `express` can be used instead of `WinFeatures.express` as this name is unique in our example.

5. A constraint is a Boolean predicate, so the sets yielded by clafer expressions and navigation need to be turned into Boolean values. This can be done using containment (`in`) between sets, equality tests (`=`), and quantifiers (see below).

6. The simplest quantifiers are `one`, `lone`, `some`, and `no` (`not` is a synonym for `no`). In the simple form, they all can be followed by a set expression. For an expression $x$, `one` $x$ means that the set specified by $x$ must be a singleton; `lone` $x$ means that the set yield by $x$ must be empty or a singleton; `some` $x$ means that the set yield by $x$ must not be empty; `no` $x$ means that the set yield by $x$ must be empty.

7. When a clafer name is used in an expression, in a position where a Boolean is expected, it is automatically cast to Boolean using the `some` quantifier. This gives a particularly natural interpretation for clafers with cardinality `0..1`. For instance, `[ express ]` really means `[ some express ]`, which is true if `express` is instantiated and false if it is not. So clafers with cardinality `0..1` behave like Boolean variables, and constraints involving them can be almost always read as Boolean conditions.

8. The quantifiers also come in a set comprehension form, where a name can be bound to elements of a set, and a predicate on the elements can be enforced. Details are further explained in section 5, using the example of `all`—the universal quantifier enforcing that all set elements satisfy a property.

9. Besides the above Clafer supports the usual set of Boolean connectives, including implication (`=>`) and bi-implication (`<=>`).

In the following section, we will be extending this constraint language with behavioral operators, starting with adding temporal logics operators, and continuing with syntactic sugar that mimics automata-like transitions.

## 4   Modeling Behaviour with Variability

Having established the key behavioral modeling practices for Clafer, we will now move towards modeling behavior. We start with discussing the notion of state space (and modeling structured state spaces), then we continue to modeling with state ma-

**Clafer: Lightweight Modeling of Structure, Behaviour, and Variability**

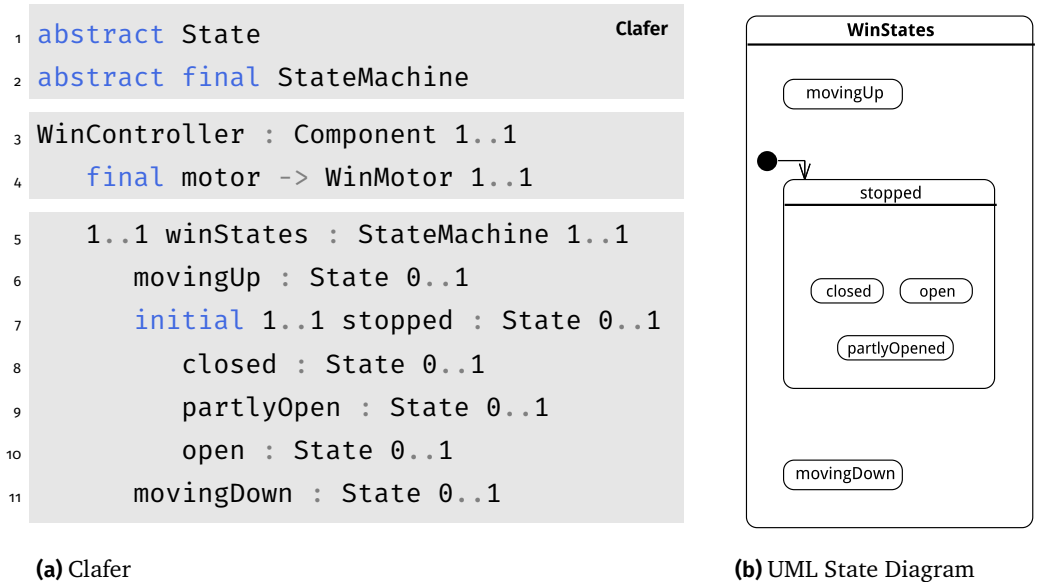

(a) Clafer  (b) UML State Diagram

**Figure 3** Hierarchical decomposition of states in the controller model

chines and scenarios. We include demonstration of several different patterns allowing introducing variability into behavioral models.

### 4.1 Hierarchical Decomposition of States as a Modeling Pattern

The need for hierarchical decomposition of states was originally recognized in the seminal work of Harel on statecharts [36], concurrently to similar developments in process algebras CCS and CSP. UML has adopted Harel's design for state diagrams and essentially also for activity diagrams. The idea has spread widely, including commercial tools, like MATLAB Stateflow.

Let's consider modeling hierarchical state decomposition in Clafer. We want to specify discrete control states for the window controller (see figure 3). In Clafer (lines 5–11), the state hierarchy is modeled using the principal nesting construct of clafers, exactly in the same way as the containment hierarchy and feature hierarchy have been modeled before. We introduce abstract clafers to denote states, state machines and WinController (lines 1–3) and use subtyping to denote the function of clafers in the model. The reference clafer `motor -> WinMotor`, is `final` meaning that the motor is always present in a controller and cannot disappear. The clafer representing a state machine, `winStates`, is not marked as `final` because it inherits that property from `StateMachine`. We always have exactly one state machine (clafer multiplicity to the right of the declaration) and exactly one instance of its children is always present (`1..1` to the left, expressing xor clafer group cardinality).[11] This effectively means that the three nested clafers (`movingUp`, `stopped`, `movingDown`) are exclusive in any snapshot, which in statechart terminology means that they are sequentially composed. The clafer `stopped` (state) is further decomposed using the same principle, into states `closed`, `partlyOpen`, and `open`. These states will store the controller's memory of what the state of the glass panel is, whenever it stops and figure 3b shows the same decomposition as a UML state diagram.

---

[11] Incidentally, this double cardinality annotation may appear confusing. In practice, many of constraints are omitted thanks to conveniently chosen default constraints. It rarely happens that both clafer (right) and group (left) cardinality need to be provided at the same time.

A model containing non-final clafers (like the above) is interpreted as an invariant over all instances, so it generates a universal language of sequences of structures satisfying this invariant. The `initial` keyword (line 7) constraints this language slightly, by stating that whenever an instance of the machine `winStates` is created, an instance of its initial child (`stopped`) will be created. In this example, this means that the first snapshot of every trace will contain an instance of `stopped`. Note that none of the children clafers of `stopped` is marked initial, which introduces non-determinism into the model: traces can start with any one of them being instantiated in the first snapshot.

Intervals can appear before and after clafer declarations. An interval constraint placed to the right of a declaration indicates *clafer multiplicity* and restricts possible cardinality of the set of instances of the declared clafer per instance of its parent clafer. So `a..b` means that each parent of the declarated clafer can have at least `a` and at most `b` instances of it. An interval placed to the left of a clafer name indicates *group cardinality*, which restricts the number of concurrently present instances of the children of that clafer.

Note that states and state machines are not first class citizens in Clafer. We use abstract types to introduce states (`State`) and state machines (`StateMachine`) in Clafer, and nesting to represent a state hierarchy. Sequential decomposition of states is realized using group cardinality constraints (`1..1 winStates` in figure 3). Parallel decomposition is realized by placing two or more clafers at the same nesting depth and not making them exclusive (for `n` concurrent regions use group cardinality `n..n` instead of `1..1`). Finally, the initial state at every level of nesting can be indicated using the `initial` keyword (for instance `initial stopped` in figure 3), which the compiler expands to a suitable temporal constraint.

The pattern for modeling state hierarchies, can be combined with the pattern for modeling messages, mentioned in the previous section. In listing 2, we recall the definitions of the clafers representing window's controller (lines 1–12) and the window's electric motor (lines 13–14). In the bottom we model several concrete commands, which can be sent to the port `WinMotor.cmd`.

In the controller automaton, we introduce constraints nested under states to enforce suitable commands modeling actuator signals in each state (see lines 5, 7, and 12). These constraints effectively are state invariants. Recall that constraints nested under a clafer are only enforced whenever an instance of this clafer exists (or more precisely

---

In particular, in this example the right-most constraint could be omitted, as `1..1` is derived by the default mechanism here. We strive to avoid using defaults and conventions in the examples as we assume that the reader is not a regular user of Clafer and would find it confusing.

**Listing 2** Do-activities in Clafer

```
1  WinController : Component 1..1
2    final motor -> WinMotor 1..1
3    1..1 winStates : StateMachine 1..1
4      movingUp : State 0..1
5        [ motor.cmd=motorUp   ]
6      initial 1..1 stopped : State 0..1
7        [ motor.cmd=motorStop ]
8      closed : State 0..1
9      partlyOpen : State 0..1
10     open : State 0..1
11     movingDown : State 0..1
12       [ motor.cmd=motorDown ]
13 WinMotor : Component 1..1
14   cmd : Port -> Command 1..1
15 abstract final Command
16 motorUp    : Command 1..1
17 motorStop  : Command 1..1
18 motorDown  : Command 1..1
```

for each such instance). Thus the motor command port is set to `motorUp` when in the state `movingUp`, and similarly for the states `stopped` and `movingDown`. This way of linking states to their semantics, is reminiscent of do-activities in UML state diagrams, yet no special support is required in Clafer (of course, this is not a precise encoding of do-activities, which have much richer semantics in the UML).

**Initial states vs default values** The curious reader may inquire why we have opted for the mechanism of initial sub-clafers, instead of providing a more general mechanism for default values, that could also be used for, say, targets of references. This is unexpected in light of our otherwise rather parsimonious attitude to language design though. In fact, we did experiment with default values in Clafer. The key difference between an initial and a default value is that an initial value is enforced strictly. If another constraint implies that another clafer is instantiated (not the one marked as initial), then an inconsistency (an error) arises, and is reported by tools. The usability expectations for default values are different: a default value should be enforced *unless* another constraint prefers instantiating another clafer. This "unless" condition is difficult to assure in a constraint language, unlike in a usual programming language. It essentially would require adding a soft-constraint facility to a language that is, so far, based on hard constraints. Note that the enforcement of another initial value than the default can be indirect, as a consequence of conjunction of several constraints. Thus, it cannot be detected syntactically or by the type system alone like in programming languages. So far, we have been unable to design a semantics that makes sense in many scenarios, is clear for users, and is implementable. Hence the compromise with `initial`. Of course, the problem of defaults remains an interesting question for future work.

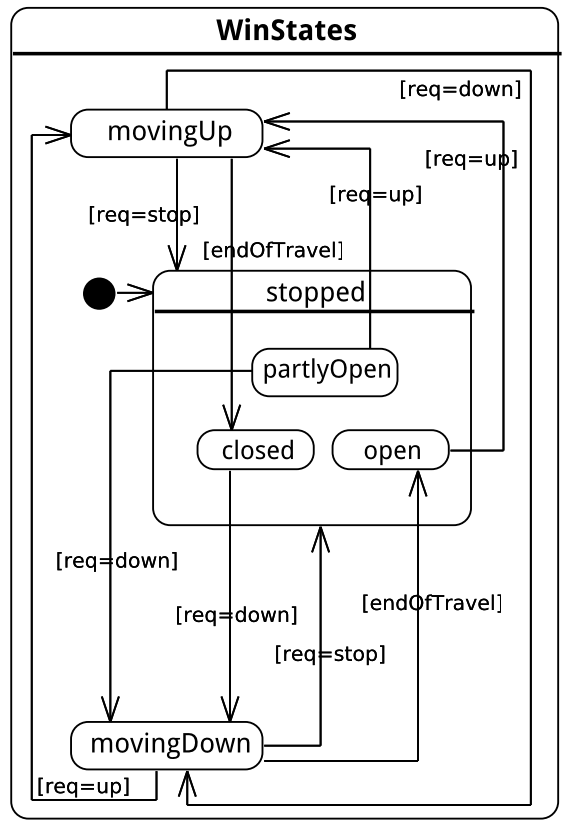

**Figure 4** UML state diagram for a control automaton for the power window controller.

### 4.2 Behaviour in Automata Style as a Modeling Pattern

Let us enrich our state model with transition constraints to control the dynamics of the underlying state machine. Assume that `UserRequest` is an abstract final type of user inputs, and the following concrete inputs are declared using subtyping (right):

```clafer
abstract final UserRequest
stop : UserRequest 1..1
up   : UserRequest 1..1
down : UserRequest 1..1
```

Moreover, listing 3 adds behaviour to the state hierarchy of figure 3a. In lines 2–3, we declare a port to receive user requests and an additional signal, `endOfTravel`, from the motor that informs the controller whenever the window cannot move any further,

■ **Listing 3** Clafer code for a control automaton for the power window controller.

```Clafer
WinController : Component 1..1
  req : Port -> UserRequest 0..1
  endOfTravel : Port 0..1
  [ lone (req ++ endOfTravel) ]
  [ always req=stop between req=down and req=up]
  [ always req=stop between req=up and req=down]
  [ always req=up between req=stop and req=expressUp ]
  [ always req=expressUp => next no req]
  1..1 winStates : StateMachine 1..1
    movingUp : State 0..1
      [ req=down --> movingDown ]
      [ endOfTravel --> closed ]
      [ req=stop --> stopped   ]
    initial 1..1 stopped : State 0..1
      closed : State 0..1
        [ req=down --> movingDown ]
      partlyOpen : State 0..1
        [ req=up --> movingUp    ]
        [ req=down --> movingDown ]
      open : State 0..1
        [ req=up --> movingUp ]
    movingDown : State 0..1
      [ req=up --> movingUp    ]
      [ endOfTravel --> open ]
      [ req=stop --> stopped ]
```

being already fully opened or fully closed. The constraint in line 4 says that at most one event can be present at a time, so either a request is present or the `endOfTravel` signal arrives (like in Alloy, `++` is a union operator and the quantifier `lone` restricts the size of the following set to at most one). This constraint allows us to restrict the Clafer semantics to the usual semantics of automata, when one event arrives at a time. In lines 5–8, we define constraints to specify sequence of user requests that can happen in the controller. The lines 5–6 specify that `stop` user request can only occur in between requests `down` and `up` or `up` and `down`. Similarly, in case if the request is `expressUp` then no request can happen afterwards (line 8). The constraints on lines 5–8 are expressed using syntactic sugar that follows Dwyer's property specification patterns [25] (see below).

New constraints have been placed under the clafers representing states. We use the operator `-->` (not to be confused with reference arrow `->` nor with implication `=>` which has no inherent temporal interpretation) to represent a temporal constraint between two predicates. If the condition on the left of `-->` holds in current state, then the condition on the right holds in the next state. So the constraint in line 11 reads: if the user request is `down` then in the next snapshot the global state will contain `movingDown`. The constraint is enforced whenever the clafer `movingUp` exists (the state `movingUp` is active). In general, the syntax of the operator $x$ `-->` $y$ where both $x$ and $y$ are Boolean constraints, respectively a precondition and a postcondition for the state change. The meaning of $x$ `-->` $y$ is that if $x$ holds in a state then $y$ must hold in the next state. We show how the operator is expanded to LTL's standard operators in section 6.

The state diagram in figure 4 represents the same behaviour as the Clafer model. Let us draw our attention that besides a regular automaton transition (like the one in line 7), we have cases of transitions that are nondeterministic (line 9) and transitions that cross state hierarchy boundaries (e.g., line 7 and line 13).

Let us summarize the pattern used for modeling automata. Like previously, events are modeled as appearing and disappearing instances clafers. They can appear simultaneously, giving a (limited) sense of true concurrency. If this is undesirable, we constrain them to be exclusive, obtaining a more standard, automata-like, semantics. Transitions are modeled using stylized temporal constraints (refer to section 6 for a complete catalog). Guard conditions can be conjoined to pre-conditions in transition constraints. Similarly outputs can be enforced by conjoining them to post-conditions (not shown), when modeled in a manner similar to do-activities shown earlier. Clafer as a language is unaware of state machines, transitions or events. Still, a modest amount of syntactic sugar and a suitable selection of abstract types make the model appear like a state machine, and the semantics act accordingly.

**Use of property specification patterns in Clafer**  Dwyer et al. have analyzed a corpus of correctness specifications written in temporal logics. They found that the operator selection and syntax design in temporal logics align poorly with what needs to be

■ **Listing 4** The express functionality.

```Clafer
expressUp : UserRequest 0..1
1..1 movingUp : State 0..1
   [ req=down --> movingDown ]
   [ endOfTravel --> closed ]
   [ req=up --> movingUp ]
   [ req=expressUp --> movingUpX ]
   initial basic : State 0..1
      [ req=stop --> stopped ]
   movingUpX : State 0..1
```

specified in practice. They extracted a collection of patterns[12] that occur repeatedly in these specifications, some of them very complex. The patterns have been named and given high level syntax (like `always ... between ... and ...`). These patterns allow writing specifications more concisely and in an understandable fashion. The patterns can be directly translated to temporal constraints by the language compiler. Inspired by this highly influential work, and extensive experience of others that the patterns are indeed useful, we have incorporated the patterns directly into Clafer. This is just a syntactic sugar exercise, as Clafer's semantics is implemented in LTL (see section 5). As can be seen in the above example, the syntax of patterns aligns very well with the syntax of other constraints. We delegate the interested reader to the original paper of Dwyer et al. and to their online collection of patterns for more details, including mappings to the temporal logics.

### 4.3 Superimposed Variability Patterns in Behavioural Modeling

Variability in product line models influences not only the selection of components, but also particular functional behaviours. We now consider superimposed-variability patterns [18] to achieve control over behaviour driven by selection of features. The superimposed variability (as opposed to modular variability) means that all variants of a model (system) are contained in a single amalgamated description. Probably the earliest form of superimposed variability was the use of conditional compilation in C, with the help of pre-processor macros. Superimposition is most-often achieved using some form of presence conditions [18] attached to model elements (here to clafers).

We first consider the *alphabet variability* pattern [48]. The idea is to control the variability in behaviour by placing presence conditions on communication channels, enforcing or forbidding certain communication depending on features. For example the express feature enables or disables the `expressUp` request. In general, if we see the model as an automaton over large possibly unbounded state-space, we are manipulating the alphabet of this automaton by selecting features [48].

---

[12] The specification patterns for LTL can be found at http://patterns.projects.cs.ksu.edu/documentation/patterns/ltl.shtml, last accessed on 2018-07-19.

We shall implement the express feature introduced in figure 1. We first introduce an optional user request expressUp (listing 4, line 1). When the feature express is present in the system, a user request expressUp causes the window to close entirely (unless the down button is pressed). We model this new behaviour by refining the state movingUp into two sub-states: basic and movingUpX, see listing 4.

When the window is movingUp it can still switch to the movingDown mode as previously, when the down request is received (line 3). Similarly it stops in the closed mode, when endOfTravel is detected. If the request received is up in movingUp state (line 5), it will still continue in movingUp state, which indicates totality of transitions. When the moving up movement is activated the machine initially enters the basic mode. In this mode, it is possible to stop, as it was possible previously (line. 8). When the request expressUp is received, the controller enters the movingUpX state, in which the stop commands are ignored until the window closes or a down request is received.

The new behaviour shall only be available, if the feature express is present. The most basic and the coarsest way to control the behaviour is to limit the existence of the request expressUp to systems in which this feature is implemented [70]:

```
[ one WinFeatures.express <=> one expressUp ]                    Clafer
```

In more refined scenarios, it might be necessary to introduce variation not just at the level of alphabet, but for individual transitions, in the style of feature transition systems [16]. We realize this by including presence conditions on static (final) features in transition guards, the so called *transition variability*. For instance the transition from line 7 could have been written as follows, to ensure that it only executes when the express feature is selected:

```
[ basic -[express && req=expressUp]->  movingUpX ]               Clafer
```

The transition variability pattern allows for finer control of behaviour than alphabet variability. In both cases, we have followed the same principle for relating feature models to behaviour, by simply using constraints. Observe that Clafer has no first-class constructs aiming at mapping features to behaviours as seen in other languages, yet modeling these mappings is direct and concise. Moreover we obtain a direct semantic integration of feature modeling, structural modeling and variability modeling as all these aspects are parts of the same model expressed in the same language. This is in contrast to orthogonal variability modeling languages (like CVL), which only relate variability to syntax of the base language, and unlike CVL and UML, which both express these aspects in several notations.

### 4.4 Architectural Variability Patterns in Behavioural Modeling

Traditionally in object-oriented systems, variability is not realized using presence conditions (or not only using presence conditions), but using suitable architectural mechanisms, including inheritance and design patterns. These are also the primary

**Clafer: Lightweight Modeling of Structure, Behaviour, and Variability**

▪ **Listing 5** Using inheritance

```
1 WinCtrWithContChime : WinController 0..1
2    chime : Port 0..1
3    [ movingUpX <=> chime ]
4 WinCtrWithChime : WinController 0..1
5    chime : Port 0..1
6    [ no movingUpX => no chime ]
7    [ chime --> no chime ]
8    [ (no chime && movingUpX) --> (movingUpX => chime) ]
```

▪ **Listing 6** Using a strategy pattern

```
1  abstract ChimingStrategy
2     active : State 0..1
3        audible : State 0..1
4  NoChiming : ChimingStrategy 1..1
5     [ not audible ]
6  Continuous : ChimingStrategy 1..1
7     [ active => audible ]
8  Intermittent : ChimingStrategy 1..1
9     [ audible --> not audible ]
10    [ not audible --> (active => audible) ]
11 WinControllerStrategy : WinController 0..1
12    chime : Port 0..1
13    strategy -> ChimingStrategy 1..1
14    [ strategy.active <=> movingUpX ]
15    [ strategy.active.audible <=> chime ]
```

variability mechanisms in architectural design languages, including AADL and EAST-ADL. In this section, we show how these mechanisms can be exploited for modeling variability for functional behaviour in Clafer.

We add a new (hypothetical) functionality to our system, namely, that whenever the window is being closed in the automatic express mode, a chime signal is being emitted to warn the users. We will consider two versions of this functionality: one when chiming is continuous (just a long beep), and one when chiming is pulsing. We will now realize this using an *inheritance-based-variability* pattern. The two variants of the functionality will be modeled in two different refinements of the window controller clafer. The proposal is presented in listing 5. In lines 1–3 we model the first functionality.

We add the (output) port `chime` to the extended controller, and a constraint stating that this port is active whenever the window is in the state `movingUpX`. The second functionality, pulsing, is modeled in lines 4–9. Again the port for controlling the sound is added, followed by three constraints. The first states that when the window is not in the state `movingUpX`, chiming is inactive. The second states that if `chime` is active now, then it is inactive in the next state (creating the pulsing effect). The third constraint says that if the chime port is inactive and we are in state `movingUpX`, then chiming should become active in the next state, but only if we remain in the state `movingUpX`.

Since our system only has one controller (see figure 2a, line 4) and both its refinements are optional, we have three possible variants of the system: without chime, with continuous chime, and with pulsing chime. We could relate these variants to a suitable sub-feature of `express` using a constraint, as shown before.

Similarly, listing 6, shows an instance of the *strategy* pattern implementing the same variability. We first define an abstract type for chiming strategies. Each of the strategies will have two state variables, one modeling when the sound is audible, and the other when the strategy is active. By simply nesting `audible` under `active`, we ensure that the strategy can only be audible when it is active (an instance of a clafer cannot exist without an instance of its parent clafer). Then we define three strategies for chiming. The first one, corresponding to the lack of chiming, like in the original `WinController`, basically states that the sound is never audible. The second one, `Continous`, is audible always when active. The third one, `Intermittent`, alternates between audible and inaudible, like `WinCtrWithChime` in the previous example. In the bottom of the figure, we show how the `WinController` is extended to interface to the strategies. We add a `chime` port, but now followed also by a reference to the current chiming strategy (line 13). The last two lines link the strategy interface to the controller interface: the first stating that the strategy is active, whenever the window is `movingUpX` (line 14), the second stating that chiming port is activated whenever the strategy decides that the sound should be audible (line 15). Like previously, choice of the strategy could be linked to a feature using constraints; which we elide.

### 4.5 Scenario Modeling Using a Pattern Based on Assertions

Importance of modeling scenarios of system executions is widely recognized. In UML, scenarios are expressed using sequence diagrams. Scenarios have many applications: they can be used to explicate user requirements (as in behaviour-driven design), or to express test executions on the system, or simply as examples leading to understanding systems functionality [6], finally they can be used to verify systems for simple liveness, safety and reachability properties.

In Clafer, scenarios are modeled as assertion constraints that are checked for consistency with the rest of the model, as opposed to being enforced on all executions of the model. This way a very strong constraint can be checked, for instance that a state is reachable, without limiting all executions to satisfy this constraint (necessarily reach this state). We use assertions like tests (to check if things work as foreseen) and constraints as specifications (to say how things must behave). For instance the following scenario (assertion) says that it is possible in our model to go from `closed` to `open` state via the `partlyOpen` state:

```
assert [ sometime closed -->> partlyOpen -->> open ]
```
<div align="right">Clafer</div>

The `sometime` quantifier is taken from Dwyer's property patterns [25]. The double-head arrow, is another transition constraint in Clafer, similar to `-->`, but not enforcing that the move will happen in the next time epoch, only that it will eventually happen. The above assertion holds (evaluates to true) if there exists an execution from the initial state that reaches a state where `closed` holds, and then in the future another state where `partlyOpen` holds, and eventually a state where `open` holds. If we omitted the `assert` keyword, and interpreted the same expression as a constraint, we would limit all executions of the model to go through the states—effectively removing executions where the window remains closed indefinitely from the semantics of the model.

Since scenarios in Clafer are just assertions admitting the entire constraint language, more complex consistencies can be checked. Negative scenarios are often useful for expressing simple safety properties. In case of the window for instance, we can check whether it is impossible that the user requests to open the window, and the window remains closed in the next step:

```
assert [ never req = down --> closed ]
```
<div align="right">Clafer</div>

Or a simple safety invariant that it is not possible for the window to be closed and continue sending the move up command (which would burn the motor).

```
assert [ never closed && cmd=motorUp ]
```
<div align="right">Clafer</div>

Both positive and negative scenarios (along with tools that can check them against the model) are extremely instrumental in debugging models. Observe that since Clafer contains LTL in its language of constraints, scenario checking contains model-checking of LTL on Clafer models. To model check property $\varphi$, add the following assertion to the model: `assert [ never ¬`$\varphi$` ]`.

### 4.6 Summary

In this section, we demonstrated various modeling idioms that can be used to model structure and behaviour of systems with variability. In addition to rich structural modeling capabilities as demonstrated before [4, 54, 62, 63], the bahaviour can be modeled as automata, as a set of properties, and as a set of scenarios, or all of them combined. We demonstrated alphabet variability, transition variability, and variability using inheritance and strategy patterns.

All of these modeling styles and idioms are supported by the basic yet universal modeling capabilities of Clafer: abstract, concrete, initial, and final clafers, nesting, inheritance, references, clafer multiplicity and group cardinality, top-level and nested constraints, assertions. We enriched the constraint language with temporal operators, property patterns, and syntactic sugar for unguarded and guarded "next-step" and "eventually" transitions.

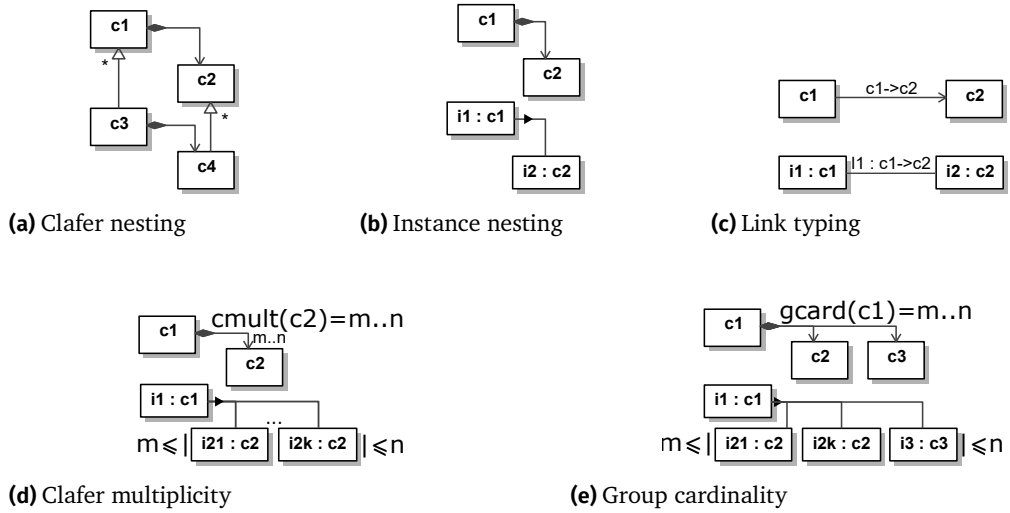

**Figure 5** Informal illustrations of selected semantic rules

## 5 The Formal Core of Clafer

We now define the core sub-language of Clafer. The core sub-language is not a technical component of the language, but an abstract methodological device—we select it from full Clafer purely for the purpose of this paper, as formally defining the entire language is impossible in a research paper (it would be too long and too boring for the reader). We carefully select a part of the language that has all the key ingredients allowing to explain how structure, behaviour, and the constraint language are interacting in the model execution. For the sake of brevity, we will assume that all clafers have super types, all constraints are written at the top level and use only basic first order and LTL operators. A more informal discussion of other aspects of the language can be found in section 6.

Let $\mathbb{C}$ be an infinite universe of discrete entities called clafers, and let **Sing** $\in \mathbb{C}$ be a designated clafer. **Sing**, for single or singleton, will serve as a default, invisible in concrete syntax, root of all models, akin to document root in many other modeling languages. We comment on the meaning of the individual points under the definition. The asterisk in superscript of a binary relation denotes a reflexive transitive closure of the relation (for instance $\twoheadrightarrow^*$ below).

**Definition 1 (Clafer Model)** *A tuple* $\mathcal{M} = (\mathsf{C}, \mathsf{A}, \rightarrow, \mathsf{cmult}, \mathsf{gcard}, \twoheadrightarrow, \blacklozenge\!\!\rightarrow, \mathsf{constr})$ *is a Clafer model, where*

1. $\mathsf{C} \subseteq \mathbb{C}$ *is a set of clafers and* **Sing** $\in \mathsf{C}$,
2. $\mathsf{A} \subseteq \mathsf{C}$ *is a set of abstract clafers, and* **Sing** $\notin \mathsf{A}$*; clafers in* $\mathsf{C}\setminus\mathsf{A}$ *are concrete.*
3. $(\rightarrow) : \mathsf{C} \hookrightarrow \mathsf{C}$ *is a partial function specifying references,*
4. $(\blacklozenge\!\!\rightarrow) : \mathsf{C} \setminus \{\mathbf{Sing}\} \rightarrow \mathsf{C}$ *is a total mapping from clafers to parents, defining a tree rooted in* **Sing***. We write* $c_1 \blacklozenge\!\!\rightarrow c_2$ *if* $(\blacklozenge\!\!\rightarrow)(c_2) = c_1$,
5. $(\twoheadrightarrow) \subseteq (\mathsf{A} \times \mathsf{A}) \cup ((\mathsf{C} \setminus \mathsf{A}) \times \mathsf{C})$ *is an* acyclic *generalization relation,*

**Clafer: Lightweight Modeling of Structure, Behaviour, and Variability**

6. *If super type is nested, then it must be nested under super type of parent: for all clafers $c_1, c_2, c_3, c_4$ if $c_2 \neq \mathbf{Sing} \wedge c_4 \twoheadrightarrow^* c_2$ and $c_3 \leftrightarrowtail c_4 \wedge c_1 \leftrightarrowtail c_2$ then $c_3 \twoheadrightarrow^* c_1$ (figure 5a),*
7. $\mathsf{cmult} \colon C \mapsto 2^{\mathbb{N}}$ *maps clafers to multiplicity constraints and* $\mathsf{cmult}(\mathbf{Sing}) = \{1\}$,
8. $\mathsf{gcard} \colon C \mapsto 2^{\mathbb{N}}$ *is a total mapping to group cardinality constraints;* $\mathsf{gcard}(\mathbf{Sing}) = \mathbb{N}$,
9. $\mathsf{constr} \subseteq \Psi[C]$ *is a set of global constraints.*

The first point of the above definition establishes a universe of clafers (simple type tags), with a distinguished root **Sing**. All other clafers will be reachable from **Sing** by navigation. We distinguish abstract and concrete clafers, analogically to abstract and concrete classes. **Sing** is required to be concrete (point 2). The $\rightarrow$ arrow represents the reference relation (point 3). It is partial because not all clafers are required to have the reference arrow. References relate clafers to clafers; analogically to how in object-oriented languages, say Ecore, references relate classes to classes. The $\leftrightarrowtail$ arrow (point 4) maps clafers to parent clafers containing them. Since **Sing** has no parent, the map is not defined for it, but it is total on the complement of **Sing**. This relation is required to be acyclic to a form a single tree, like a syntax tree of a program, or an element tree of a model.

The $\twoheadrightarrow$ arrow (point 5) is the relation defining the generalization relation (a transposition of the inheritance relation). Abstract clafers are only allowed to be generalized by abstract clafers, while concrete clafers can be generalized by either abstract or concrete clafers. This is stricter than in, for instance, Java where abstract classes can inherit from concrete classes. The difference is primarily caused by lack of operations in Clafer, so there is no way to add an element of abstraction to a concrete clafer by extending it. This restriction does not appear fundamental though, and it might be lifted in the future. Point 6 captures the condition allowing simultaneous covariant refinement/generalization of nested clafers and their parents. See figure 5a for a visualization of this property. Contravariant refinement of nested clafers is presently not supported.

The final part of definition associates constraints with clafers: clafer multiplicity constraints (point 7), group cardinality constraints (point 8) and global constraints (point 9). In the abstract formalization, the clafer and group constraints are modeled as simple subsets of natural numbers, with the natural correspondence to the interval syntax used in the previous sections: an interval of naturals $[a; b] \cap \mathbb{N}$ represents the constraint a..b. **Sing** will only have a single instance, so its clafer cardinality is 1. As **Sing** contains all top-level clafers, its group cardinality is all natural numbers—this allows to capture models of arbitrary size in the semantics. The careful reader has noticed that the global constraints are indeed global (point 9), not nested under clafers like in preceding sections. This yields no loss of generality. In section 6 we explain how nested constraints can be translated to global constraints without changing the intended semantics. The global constraints (point 9) $\Psi[C]$ are written using the following grammar:

$$\psi ::= \mathtt{true} \mid \mathtt{all}\, x : e | \psi \mid \mathtt{let}\, x : e | \psi \mid e\, \mathtt{in}\, e \mid \mathtt{not}\, \psi \mid \psi \oplus \psi \mid \mathsf{X}\, \psi \mid \psi_1\, \mathsf{U}\, \psi_2$$
$$e ::= c \mid x \mid e.e \mid e.\mathsf{dref} \mid e.\mathsf{parent}$$

where $c$ ranges over clafers in $\mathbb{C}$ and $x$ over variables names (not in $\mathbb{C}$). Constraints $\psi$ are build of: the literal true, universal quantification, let bindings, set inclusion, negation, propositional connectives ⊕, and temporal path expressions of the linear temporal logics (LTL). Expressions (e) consist of clafer names, variables references, navigation expressions, dereferencing expression, and parent navigation.

The dereferencing expression (e.dref) follows the reference arrow from all clafer instances resulting from evaluating e. This yields the set of the target instances of reference arrows. The parent navigation expression (e.parent) follows the containment arrow ↢ upwards, so given a set of clafer instances resulting from computing $e$ gives a set of clafer instances that are their parents in the containment hierarchy.

In the following, we assume that all identifiers in a model are unique, all variables in constraints must be bound in quantifier expressions or in let-bindings. These are not fundamental restrictions, but only introduced here for simplicity of presentation.

Let $\mathbb{I}$ be an infinite universe of clafer instances with a designated instance sing $\in \mathbb{I}$ and $\Pi$ be the infinite set of model instances. We now define a model instance:

**Definition 2 (Model Instance)** *A model instance $\pi$ is a tuple* $(I, L, \blacktriangleright)$ *where*
- $I \subseteq \mathbb{I}$ *is a set of clafer instances with* sing $\in I$
- $L : I \hookrightarrow I$ *is a partial function representing links between instances,*
- $(\blacktriangleright) : I \setminus \{\text{sing}\} \to I$ *is a total mapping from instances to parents, which defines a tree rooted in* sing. *We write* $i_1 \blacktriangleright i_2$ *if $i_1$ is the parent of $i_2$, so* $(\blacktriangleright)(i_2) = i_1$.

Types in Clafer are time-independent: same instance is always typed by the same clafer (static typing). The instance typing relation ity is a subset of $\mathbb{I} \times \mathbb{C}$ with ity(sing, **Sing**): ity links a clafer instance with its type (a clafer), and in particular it links sing with **Sing**. Given a model, we generalize typing of instances by composing it with the reflexive transitive closure of the generalization relation ($\twoheadrightarrow^*$):

$$\text{ity}^* ::= (\twoheadrightarrow^*) \circ (\text{ity}) \subseteq \mathbb{I} \times \mathbb{C}$$

Basically, the ity* relation links an instance not only to its direct clafer, but also all clafers that generalize it in the inheritance hierarchy. As a result the ity relation is a function (maps an instance to a single value), while ity* is multivalued if inheritance is used.

We shall now specify what does it mean for an instance to satisfy a model (we comment on the key aspects of the definition subsequently):

**Definition 3 (Structural Satisfaction)** *Given an instance $\pi$ and a model $\mathcal{M}$ defined as above, we say that $\pi$ satisfies $\mathcal{M}$, $\pi \models \mathcal{M}$, if the following hold:*

1. *Abstract clafers have no direct instances: for $i \in I$, $c \in C$ with* ity$(i, c)$ *also $c \notin A$,*
2. *Parent relation of instances is well typed: whenever $i_1 \blacktriangleright i_2$ then there exists clafers $c_1, c_2 \in C$ such that* ity$^*(i_1, c_1)$ *and* ity$^*(i_2, c_2)$ *and $c_1 \leftrightarrow c_2$ (figure 5b).*
3. *Links are well typed by references: for any link* $L(i_1) = i_2$ *there exist clafers $c_1, c_2 \in C$ such that* ity$^*(i_1, c_1)$ *and* ity$^*(i_2, c_2)$ *and $c_1 \to c_2$.*
4. *Link map is locally injective: for all clafers $c_1 \leftrightarrow c_2$ and instances $i_1, i_2, i_3 \in I$ if* ity$^*(i_1, c_1)$, ity$^*(i_2, c_2)$, ity$^*(i_3, c_2)$ *and $i_1 \blacktriangleright i_2$, $i_1 \blacktriangleright i_3$ then* $L(i_2) \neq L(i_3)$ *(figure 5c).*

**Clafer: Lightweight Modeling of Structure, Behaviour, and Variability**

5. *Every instance satisfies its clafer's multiplicity constraint: for every pair of clafers $c_1$, $c_2 \in C$ such that $c_1 \leftrightarrow c_2$ and for every clafer instance $i_1 \in I$ such that $\text{ity}^*(i_1, c_1)$ we have that $|\{i_2 \mid i_1 \blacktriangleright i_2 \text{ and } \text{ity}^*(i_2, c_2)\}| \in \text{cmult}(c_2)$, (figure 5d).*
6. *Children satisfy parent's group constraint: for each instance $i_1 \in I$, clafer $c_1 \in C$ if $\text{ity}^*(i_1, c_1)$ have $|\{i_2 \in I \mid \exists c_2. c_1 \leftrightarrow c_2 \land \text{ity}^*(i_2, c_2) \land i_1 \blacktriangleright i_2\}| \in \text{gcard}(c_1)$ (figure 5e).*

The definition largely follows semantic models for class diagrams (see for instance Fahrenberg, Acher, Legay, and Wąsowski [26]), with few Clafer specific aspects. We first (item 1) ensure that abstract clafers (like abstract classes) cannot be instantiated directly. Specifically, the relation $\text{ity}(c, \cdot)$ denotes all direct instances of clafer $c$ without instances of inheriting types. Second, there must exist a type mapping from links in the instance graph into references in the type graph, preserving types of endpoints (item 2–item 3). The main difference from class diagrams is that all links of an instance are a *single set* in Clafer (L($i$) )—they are directly identified by an instance, unlike in class diagrams where multiple link sets may be held in a link, so they are generally maps from association names. The item 5th item forces links to follow multiplicity constraints—this is not substantially different from class diagrams.

The fourth item in the definition is the construct that enables removing named references from Clafer, but allows using nested clafers as reference names (reification of associations in UML terms): if two instances are nested under the same parent instance, they are not allowed to point (using L($\cdot$)) to the same target instance. This basically makes nested instances to become links for practical purposes: an instance identifies the source (the parent instance) and target it points to, like a link. Finally, the last item ensures that instances nested under a clafer instance must fulfill the cardinality constraint of its parent. This aspect is lifted from the semantics of the feature models (see a simple formulation in our prior work [21]).

We continue to give semantics to Clafer expressions (used mostly to formulate constraints). An environment $\varepsilon$ maps variables to pairs of clafer instances and model instances, [] denotes an empty environment, an $\varepsilon \dagger [x \mapsto y]$ represents an environment obtained from $\varepsilon$ by binding a variable $x$ to $y$, where $y \subseteq \mathbb{I} \times \Pi$ is a set of pairs of clafer instances and model instances. Given a Clafer model, the semantics of expression e in an environment $\varepsilon$ contains pairs, each pair $(i, \pi) \in [\![ \text{e} ]\!]_{\varepsilon, \pi}$ consists of a clafer instance and a model instance.

$$
\begin{aligned}
[\![ c ]\!]_{\varepsilon, \pi} &= \{(i, \pi) \mid i \in I_\pi \land \text{ity}^*(i, c)\} \\
[\![ x ]\!]_{\varepsilon, \pi} &= \varepsilon[x] \\
[\![ \text{e}_1.\text{e}_2 ]\!]_{\varepsilon, \pi} &= \{(i_2, \pi') \in [\![ \text{e}_2 ]\!]_{\varepsilon, \pi} \mid \exists i_1 \in I_{\pi'} \text{ such that } i_1 \blacktriangleright_{\pi'} i_2 \land (i_1, \pi') \in [\![ \text{e}_1 ]\!]_{\varepsilon, \pi}\} \\
[\![ \text{e.dref} ]\!]_{\varepsilon, \pi} &= \{(i_2, \pi') \mid \exists (i_1, \pi') \in [\![ \text{e} ]\!]_{\varepsilon, \pi} \text{ and } L_{\pi'}(i_1) = i_2\} \\
[\![ \text{e.parent} ]\!]_{\varepsilon, \pi} &= \{(i_1, \pi') \mid \exists (i_2, \pi') \in [\![ \text{e} ]\!]_{\varepsilon, \pi} \text{ and } i_1 \blacktriangleright_{\pi'} i_2\}
\end{aligned}
$$

We discuss the above definition of the expression language line-by-line. First, any clafer name $c$ (interpreted in the context of a model instance $\pi$ and expression variables environment $\varepsilon$) denotes all instances of clafer $c$ in model instance $\pi$. This is akin to OCL's `allInstances` operator, albeit using much less syntax. The clafer instances in the computed set, in this case, and all the later cases, are paired with the model instances for technical reasons (it allows to define the semantics slightly more compositionally). The reader may wonder why the semantics of expressions is given without any direct reference to the model—there is a reference indeed, although somewhat subtle: recall that the ity* semantic operator consults the model hierarchy definition to check the inheritance hierarchy.

A constraint variable $x$ (that would typically be introduced by a quantifier or a `let` expression) denotes the set bound to it in the current environment $\varepsilon$. The environments are not build in the set-expression language, but in the constraint language (below).

The dot operation is used for navigation. The navigation restricts the set of all clafer instances designated by $e_2$ to only those that are nested under instances of clafers designated under $e_2$. Thus navigation, like in Alloy, is a relational join operator that computes an image of the parent-child relation. Here, the ▶ operator makes an implicit reference to the model again, indirectly achieved by structural semantics in definition 3 requiring that instances are only nested under parents of appropriate type.

The `dref` operation allows to navigate along the reference link (as opposed to navigating down the nesting hierarchy, like the usual dot does). The `parent` name designates navigation upwards in the nest hierarchy (as opposed to the dot alone that navigates downwards).

A *trace* is an infinite sequence of model instances: $\sigma = \pi_0, \pi_1, \ldots$; we write $\sigma^j = \pi_j, \pi_{j+1}, \ldots$ for a suffix starting at $j$th position in $\sigma$, and $\sigma[k] = \pi_k$ for a specific instance ($\pi_k$) at $k$th position from the head. In particular $\sigma^1$ always denotes a tail of trace, $\sigma[0]$ always denotes the head of trace $\sigma$, and $\sigma^j[0]$ denotes the head of the suffix of $\sigma$, which starts at the $j$th position.

With the semantics of set expressions in place we can now write the semantics of constraints (that are Boolean predicates). See definition 4. The definition follows the work of Abadi [1], who has shown how to combine linear temporal logics (LTL) with first order logics. The atomic expressions in the logics are constant `true` and set predicates (membership and inclusion both achieved using the `in` operator). The predicates use the semantics of set expressions defined above. Observe that the semantics of `all` introduces a new variable $x$ into the variable environment (that we used to interpret the set expressions before). Note that other quantifiers (herunder `no`, and `some`) can be derived using negation and `all`. The special quantifiers `one` (exactly one) and `lone` (at most one) can be defined using the set cardinality operator (not shown in this section). A key difference from the standard result of Abadi, is inclusion of the `let` construct that allows to bind a name to a set of clafer instances at a particular time. This provides an easier way to name and refer to structures across time epochs than with just using first order quantifiers.

The last four cases are standard in LTL semantics definitions. In particular the last two lines define the core temporal part of clafer using the next operator (X) that constraints the configuration of instances in the next state, and the until operator (U) which restricts all the states from now until a condition is met. Their formalization is standard, as defined by Pnueli [57]. These two operators are sufficient to derive the entire LTL as syntactic sugar, and further extensions used in Clafer, such as, Dwyer's property specification patterns (for instance `always` and `never`) and transition constraints (`--> `). The ⊕ symbol stands for Boolean connectives.

**Definition 4 (Constraint satisfaction)** *The satisfaction of a constraint $\psi$ in environment $\varepsilon$ by a trace $\sigma$ (written $\sigma, \varepsilon \models \psi$) is defined by structural induction:*

| | | |
|---|---|---|
| $\sigma, \varepsilon \models \text{true}$ | always | holds for each trace $\sigma$ and environment $\varepsilon$ |
| $\sigma, \varepsilon \models \text{all } x : e \mid \psi$ | iff | $\forall \bar{x} \in [\![e]\!]_{\varepsilon, \sigma[0]}.\ \sigma, \varepsilon \dagger [x \mapsto \{(\bar{x}, \sigma[0])\}] \models \psi$ |
| $\sigma, \varepsilon \models \text{let } x = e \mid \psi$ | iff | $\sigma, \varepsilon \dagger [x \mapsto [\![e]\!]_{\varepsilon, \sigma[0]}] \models \psi$ |
| $\sigma, \varepsilon \models e_1 \text{ in } e_2$ | iff | $\{i \mid \exists \pi.(i, \pi) \in [\![e_1]\!]_{\varepsilon, \sigma[0]}\} \subseteq \{i \mid \exists.(i, \pi) \in [\![e_2]\!]_{\varepsilon, \sigma[0]}\}$ |
| $\sigma, \varepsilon \models \neg \psi_1$ | iff | $\sigma, \varepsilon \models \psi_1$ does not hold |
| $\sigma, \varepsilon \models \psi_1 \oplus \psi_2$ | iff | $(\sigma, \varepsilon \models \psi_1) \oplus (\sigma, \varepsilon \models \psi_2)$ |
| $\sigma, \varepsilon \models \text{X } \psi$ | iff | $\sigma^1, \varepsilon \models \psi$ |
| $\sigma, \varepsilon \models \psi_1 \text{ U } \psi_2$ | iff | $(\exists k \geq 0.\ \sigma^k, \varepsilon \models \psi_2) \land (\forall 0 \leq j < k.\ \sigma^j, \varepsilon \models \psi_1)$ |

In the following, the model and all model instances in a trace are as defined in definition 1–definition 2. We index constituents of the $j$th element of a trace $\sigma$ by $j$, so for instance $I_j$ denotes the set of clafer instances of the model instance $\sigma[j]$.

**Definition 5 (Trace Satisfaction)** *A trace $\sigma$ satisfies a model $\mathcal{M}$ iff*

1. *All instances satisfy the model structurally: $\forall j \geq 0. \sigma[j] \models \mathcal{M}$ as per definition 3*
2. *Instances that disappeared cannot reappear: if $i \in I_j \land i \notin I_{j+1}$ then $\forall l > j. i \notin I_l$*
3. *A clafer instance can not change parent: for each snapshot $j \geq 0$ if $i_1, i_2 \in I_j$ and $i_1 \blacktriangleright_j i_2$ then $i_1 \blacktriangleright_{j+1} i_2$ or $i_2 \notin I_{j+1}$.*
4. *All constraints must be satisfied as in definition 4: $\sigma, [] \models \psi$ for each $\psi \in \text{constr}$.*

The first point of the above definition ensures that the evolution of instances obeys the main structure of the model hierarchy, while the last point ensures that it obeys all the first order (extra) constraints. Both structure and extra constraints are invariant along the trace. The two middle items simplify the evolution of the behavior regarding clafer identity (cannot reappear) and nesting (cannot change parent).

We decided that clafer instances cannot change parents during execution not only because this simplifies the language mechanics, but also because this is a standard choice for direct aggregation in most languages. For instance, this is the case with black-diamond nesting in the UML class diagrams, or with struct and class member nesting in C++. (Incidentally, in Java, objects can change parents, because all nesting is done by references. This is consistent with our choice that reference targets can change at runtime in Clafer. Strictly speaking, in Java there is no way to nest objects, but only to nest references, and references cannot change parents in Java, again consistently with our definition of Clafer nesting.)

## 6 Design and Implementation

This section discusses the basic language constructs, syntactic sugar, and implementation details of Clafer, primarily focussing on behavioural aspects. It explains how the semantic core is used to scaffold a richer language.

**6.1 Defaults and syntactic sugar**

The models presented so far relied on a few defaults and syntactic extensions, that were not included in the core semantics of the previous section. We present all these extensions as syntactic sugar. We use the following EBNF conventions when presenting concrete syntax: terminal symbols (keywords, operators) are written in quotes (`""`), optional elements in square brackets `[ ]`, and repeating elements in curly brackets `{}`. Whenever we use indentation in a piece of grammar, we mean that an increase of indentation in concrete syntax is required in the corresponding location. We also remark that more details about structural aspects of Clafer can be found in [4].

Clafer's concrete syntax defines only two categories: clafer declarations and constraints.[13] Only the clafer name is required in a clafer declaration, the other parts are optional. All the components are listed in the following grammar scheme:[14]

```
["abstract"] ["final"] ["initial"] [gcard] name [super]         grammar scheme
        [reference] [cmult] [init]
    { nested declaration }
```

By default, a clafer is not abstract, not final, and not initial. Default group cardinality is `0..*`; the default super clafer is `clafer` (a specially designated clafer); and no type is referenced by default. The default multiplicity is `1..1` for clafers whose parent have group cardinality `0..*`, and `0..1` otherwise. Cardinality and multiplicity are specified by an interval $n..m$ or $n..*$, with keywords abbreviating commonly used cases: `xor` for `1..1`, `or` for `1..*`, and `mux` for `0..1` for groups, and ? for `0..1`, * for `0..*`, and + for `1..*` for clafer multiplicities. The super-clafer is written as colon followed by a name. The reference syntax is `-> name` followed by a name if referencing a set, and `->> name`, when referencing a bag. Initializer allows setting value of the reference to a given expression: constant `= exp` and default `:= exp`. Nested declarations are indented.

The syntax of constraints is:

```
context                                                         grammar scheme
    [ assert ]"[ " φ("this") "]"
```

where the context is either empty (for a top-level, unindented constraint) or it is a clafer declaration. The actual constraint $\varphi$ can refer to any clafer names in the model, including the special identifier `this`, which indicates an instance of the context. **Sing** is the context for toplevel Clafers.

By default a constraint is not an assertion but a part of the model. This means that the constraint limits what are legal instance (for instance for the instance generator, trace generator or configurator). This can be changed by prefixing the constraint with `assert`. Then the constraint is checked by the compiler, and an error is reported if it is inconsistent with the model. Assertions do not have to be consistent between each other. They are used to test and debug models.

---

[13] There exists the third category: *optimization objectives*, which we do not consider in this work. Static Clafer models can be multi-objectively optimized to find a Pareto front of non-dominated model instances given a set of optimization objectives [54, 62, 63].

[14] The implementations of concrete syntax uses white-space sensitive grammar in the style of Python and Haskell to make the use of braces optional.

The core language assumes all constraints to be top-level; however, in our examples we nested the constraints under clafers. Every nested constraint can be lifted into top-level by adding explicit quantification over the instances of context clafers. For example, the constraint nested under B on line 3 below is lifted one level up in line 4 and to top-level in line 5 as follows:

```
1  A 2..5                                             grammar scheme
2     B *
3       [ φ("this") ]
4     [ G(all b : this.B | φ(b/this)) ]
5  [ G( all a : this.A | G(all b : a.B | b => φ(b/this))) ]
```

Lifting a constraint a level up in the hierarchy means that it holds for all instances of its context clafer (thus the `all` quantification above) and as long as they exist, hence we add the LTL quantifier globally (`G`) above. The notation `b/this` means a substitution of all occurrences of `this` by `b`. Above, the constraint from line 3 is first lifted to the context of A (line 4) and then again to the top-level context (line 5). After this is done the top-level `this` is replaced by **Sing**, which brings us to constraints of the previous section (not using the `this` keyword). To keep the language compatible with structural Clafer, top-level constraints that have not been lifted are preceded by the LTL quantifier `G`.

The above globalization rewrite allows to derive semantics (and implementation) of Clafer from the core language presented before. In practice, the compiler does not directly use the globalization for performance reasons, but resorts to an equivalent encoding that maintains the information about nesting of constraints. This helps the performance of CSP and SAT solvers used in the back-end.

The Clafer constraint language takes major influence from Alloy [42]. It supports negation (!), conjunction (&&), and disjunction (||) of Boolean expressions. For set operators, Clafer supports set union (`++` and `,` (syntactic sugar)), difference (`--`), intersection (`**`), relational join (`.`) and subset (`in`). Set equality is de-sugared into two-way set inclusion. Borrowed from Alloy, Clafer also supports first-order logic quantifiers: `all` ($\forall$), `some` ($\exists$), `no` (none), `lone` (less than one), and `one` ($\exists!$). The quantifier `some` is added by default allowing for concisely asserting the presence of clafers: e.g., [ chime ] is de-sugared to [ `some` chime ].

Navigating in the hierarchy of clafers is performed using the join operator (`.`). In our examples, we could simply refer to clafer names by taking advantage of the compiler's built-in name resolution mechanism.[15] For example, in listing 3, line 7, the name `req` is resolved to a path `this.parent.parent.req`, and the name `movingDown` is resolved to `this.parent.movingDown`.

The behavioural language supports the basic LTL operators: `G` (globally), `F` (eventually), `X` (next), `U` (until), and `W` (weak until), all of which are obtained from the basic next (`X`) and until (`U`) in the standard way. Temporal operators `initially` and `finally` respectively indicate the first snapshot in which an instance of the context clafer has appeared and the last time moment before it disappeared. They are de-sugared as follows, before being lifted to top-level:

**schema rewrite**

```
"[ initially"φ("this")"]"  ⇝  "[ (no this && X this) => X"φ("this")"]"
"[ finally"φ("this")"]"    ⇝  "[ (this && X no this) => "φ("this")"]"
```

A condition $\varphi$ holds initially for the context clafer `this` (first rewrite, left) if whenever the `this` clafer appears, so first does not exist in a state and then exists in the following state (right), the property $\varphi$ holds in the next state (in the first epoch when `this` appears). For top-level constraints, `initially` simply prevents adding the default `G` (mentioned above). A property $\varphi$ holds finally, whenever it holds in the very last epoch in which the context clafer `this` existed (the second rewrite).

The unguarded next-step transition `P -->  R` and multi-step transition `P -->> R` are de-sugared as follows:

**schema rewrite**

```
"[ "φ"-->  "ψ"]"    ⇝  "[ "φ"=> X"ψ"]"
"[ "φ"-->> "ψ"]"    ⇝  "[ "φ"=> (this &&"φ") U"ψ"]"
```

The meaning of a strong, next-step, transition arrow (`-->`) is thus: If precondition $\varphi$ holds in a state then the post-condition $\psi$ must hold in the next (`X`) state. The obligation is expressed using the implication arrow (`=>`). The meaning of the weak, multi-step, transition arrow `-->>` is captured using the until operator: if the precondition is $\varphi$ is satisfied in a state then it must continue to be satisfied until the post-condition $\psi$ is eventually satisfied. Also the context clafer (the source clafer) has to continue to exist until the post-condition is satisfied. The `-->>` transition is inevitable, but its timing is not restricted.

After de-sugaring, the constraints are lifted up in the hierarchy as described earlier. This makes sure that they are only active when the instance of the context clafer (often representing a state) is present. Since writing guard conditions on transitions is a common pattern in modeling, an alternative syntax allows placing a condition on the arrow operator, as in: $\varphi$ `-[` $\gamma$ `]->` $\psi$. The guarded next step transition `P -[Q]-> R` and the guarded multi-step transition `P -[Q]->> R` are de-sugared as follows:

**schema rewrite**

```
"[ "φ"-["γ"]-> "ψ"]"     ⇝  "[ "φ"&&"γ"=> X"ψ"]"
"[ "φ"-["γ"]->> "ψ"]"    ⇝  "[ "φ"&&"γ"=> (this &&"φ") U"ψ"]"
```

---

[15] Name resolution is a convenience mechanism inherited from an earlier version of Clafer. We refer the reader to prior work [4] for the details, as it has no bearing on the present contribution. In short, it allows to use unqualified clafer names as long as they are unambiguous.

Note that in both cases, the meaning is equivalent to the unguarded patterns with the precondition strengthened to be $\varphi \wedge \gamma$. In Clafer, the arrow guard is mostly to be used as a readability device: the intention is to use $\varphi$ to constraint what is the legal source state configuration, and to use $\gamma$ to add additional conditions (for instance, a dependency on feature configuration).

If a clafer c is qualified as `initial`, then a constraint [ `initially` c ] is added to its parent's constraints. The modifier `final` is expanded to the following constraint:

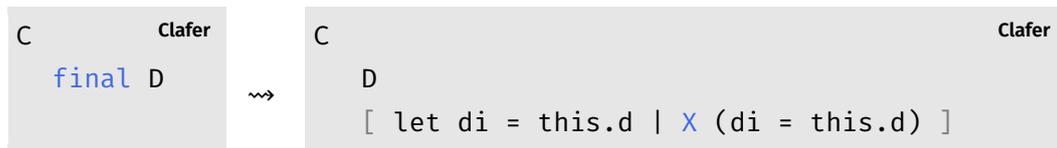

### 6.2 Implementation

The release 0.5.0 of the Clafer compiler (https://github.com/gsdlab/clafer/tree/0.5.0) is capable of compiling both static and dynamic models. For static models, the compiler generates input for two back-end reasoners: Alloy and Choco4. For dynamic models, the official compiler can output de-sugared models and generate HTML output for publishing. An experimental generator of Alloy input is also included in the release. The parser is generated from the extended grammar definition.[16] De-sugaring and HTML generation are implemented manually.

We modified the Clafer-to-Alloy compiler to enable trace generation by bounded checking of consistency of temporal constraints using Alloy Analyzer. Thus, even though the semantics of the language presented in section 5 is over unbounded traces, the implementation uses a bounded state-space. The encoding [12] is able to find finite witnesses for some temporal properties over infinite traces. We rely on an encoding proposed by Cunha [17], applying the *local state idiom*, that is, adding the time index as the last column of all mutable relations. We have extended Cunha's encoding to account for the hierarchical nature of Clafer. Cunha considers all Alloy relations as *immutable*, besides the relations in the global and local state idioms. In Clafer, on the other hand, only the top-level concrete final clafers (and their direct final descendants) are truly immutable. Final clafers nested under non-final clafers are immutable only as long as the instance of their context clafer exists, so they need to be extended with a temporal dimension like all other mutable relations. As a result of executing the Alloy's `run` command we obtain an instance which contains a single execution trace, if such exists.

---

[16] https://github.com/gsdlab/clafer/blob/0.5.0/src/clafer.cf, last accessed on 2018-07-19.

## 7 Evaluation

We conduct qualitative evaluation of behavioural Clafer by contrasting it with four modeling languages: Live Sequence Charts, SysML, AADL, and Temporal OCL. We aim to answer the following research questions:

**RQ1** *Does Clafer offer distinct advantages over other languages regarding modeling structural variability and behavioural variability?*

**RQ2** *What kind of tooling is possible for Clafer given its semantics model vs what kind of tools are possible for other modeling languages in this space?*

**RQ3** *Does Clafer offer distinct advantages to language implementers comparing to other modeling languages in this space?*

Obviously, despite a rather binary yes/no formulation of the above questions the answer to them is quite nuanced. In the analysis we will observe a number of advantages of Clafer, but also a number of weaknesses against distinct features of competing languages. Consequently, an indirect goal of this evaluation exercise is to establish a foundation in the next step of the evolution of Clafer, and to inspire designers of new modeling languages.

**Subject selection** All languages selected for comparison are general purpose modeling languages, and cover a broad range of vertical domains. They do differ widely in their semantic foundations. Initially, we identified languages that are close in nature to Clafer: they allow modeling evolution of structures (Temporal OCL), and modeling of behavioural scenarios (Live Sequence Charts). We then broadened to languages that combine structure and behaviour with some support for use case feature modeling, structural modeling of components, scenarios, trace generation (SysML, AADL).

We have deliberately not selected any modular variability modeling languages such as FORML [66] and Delta modeling [64] for this comparison. These languages take very different approach to handling variability, by maintaining modular feature descriptions—a system model is composed of modules representing features. The modular approaches are much less popular than amalgamated approaches (where Clafer, and the comparison subjects belong). While the debate of the trade-off between modular and amalgamated variability modeling languages is clearly not over, we decided against studying it in this paper. Studying the trade-off deserves a well designed study of itself, and could not be given justice as a short part in this already long paper. Importantly, any comparison with FORML and Delta modeling, would teach us more about the trade-off than about the advantages of Clafer.

There are also more prosaic reasons for this selection. To our best knowledge, FORML does not support scenario modeling. Delta modeling is a method for handling variability on top of existing languages. To cover the scope of the evaluation (structure, behavior, scenarios, constraints), we would need to develop something akin to Delta-UML and evaluate this creation. To the best of our knowledge, a Delta extension of UML does not exist (the closest option is Delta-Simulink [35], with very limited support for structural modeling using block diagrams).

**Clafer: Lightweight Modeling of Structure, Behaviour, and Variability**

**Method**  We follow the case study method, inspired by the idea of chrestomathy [65], where the same case task is solved in multiple languages. The group of authors has analyzed qualitatively the following four aspects of each of the languages used:

*Structural variability (RQ1):*  The ability of each language to model variability in the architecture, ease of integration with external variability models, and also the process of generating concrete variants.

*Behavioural variability (RQ1):*  The ability of each language to model variability in system's dynamics. We use the behavioural variability patterns (such as *alphabet* variability, *transition* variability) presented above in the paper, and evaluate the support the languages provide for each of these patterns.

*Tool support for language users (RQ2):*  We report what kinds of tools exist, and what tools are potentially possible, that would make it easier and more effective to use the language, for instance visualization, consistency checking, etc.

*Support for language implementers (RQ3):*  The aspects relevant for developers of tools for the language, such as small specification size, support for integrating disparate model views (structure, behaviour, variability), etc.

Language comparisons are prone to bias and opinion. We wanted to make the discussion possibly systematic and set the same frame of reference for all the subjects. In order to answer RQ1, we used the power window case study to create models in each of these languages, capturing aspects of structure, behaviour, and variability.[17] The modeling was performed by a student in Generative Software Development Lab in Waterloo (one of the authors), who is well experienced in modeling. We used documentation, books, and available expertise in the lab, and among authors to inform and critique the created models, always striving to maintain a positive mode (can this be modelled better in this language?) during critique sessions. For each of the used languages we had at least one person, different from the modeller, who was considered a local expert on the language (defined as a person who used the language before, in another project). In our experience, enforcing a common process and setup for all the languages, helped to reduce bias of the discussion.

Arguably, answers to questions RQ2 and RQ3 can only be given in a qualitative and approximate way, as development of language tools in a controlled experiment is not feasible for realistic languages. Yet, we believe that these are important questions to discuss. We use our experience in language design, implementation and in definitions of formal semantics to judge the potential of tooling when addressing RQ2. For RQ3 we settled on the size of the specification as the most independent reliably computable metric for a language, shared by all languages. We enrich it with discussion of the language specific implementation aspects. We are aware of the threats to validity of both of these discussions, and return to this problem in section 7.6. Of course, the discussion of tools and implementation effort for already existing languages, already implemented, and already equipped with tools is somewhat belated. However, we hope that the design decisions stand out in this analysis and can influence designers and implementers of new languages.

---

[17] Models are available from https://uwaterloo.ca/wise-lab/publications/clafer-lightweight-modeling-structure-and-behaviour, last accessed on 2018-07-19.

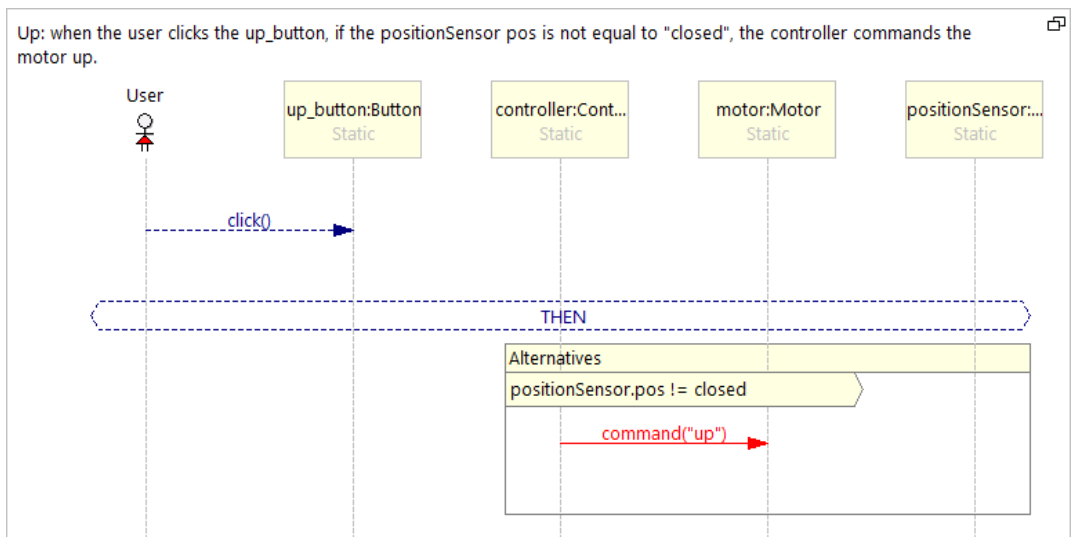

**Figure 6** Live Sequence Chart for user-up scenario

The discussion below is organized by the languages and the above dimensions of analysis. We begin with live sequence charts, then shift to SysML, AADL and Temporal OCL. For each of the languages, we compare it with Clafer along the above four dimensions. An executive summary of the comparison along these dimensions is also shown in table 1.

Independently, of this experiment we also reflect on the design of Clafer and its features in section 8, which presents a from-within analysis, as opposed to the present section which takes a from-outside perspective, comparing with other languages.

## 7.1 Live Sequence Charts

*Live Sequence Charts* (LSCs) are an extension of *Message Sequence Charts* (MSC) able to express existential and universal temporal properties [22]. LSCs have limited capabilities to model structures, and their classical use case is modeling execution traces and scenarios (both positive and negative). In contrast to Clafer, LSCs primarily help designers to specify execution behaviour, without explicit support for modeling evolution of structures over time. Although the genesis of behavioural Clafer and Live Sequence Charts seem different, both of these languages can be used to create a dynamic model with temporal constraints.

We briefly revisit main features of LSCs. To experiment with the languages, we have created the case study model using the tool PlayGo for LSC creation [38]. The built-in natural language play-in feature of the tool generates LSCs from the use case specifications. One such generated LSC is shown in figure 6. The input natural language use-case is displayed above the diagram in figure 6. The figure also shows how LSCs impart universal and existential quantifiers to traditional MSCs at each level. At the topmost level, each chart can either be quantified with universal or existential execution. Moreover, figure 6 shows a chart with universal quantification (indicated by being placed inside a solid-line box), since in this scenario, once the user clicks the `up_button`, the execution sequence is valid for *all* states of the system. Messages in the chart are also characterized as universal (message of controller sending the up command) or existential (message showing the user clicking up) through the use of solid or dotted lines, respectively.

**Structural variability (RQ1)**   The primary objective of live sequence charts was to enhance the expressive power of message sequence charts. However, since the semantics of LSCs are based on top of MSCs, they are much less expressive than a general purpose modeling language. For instance, LSCs fall short of capturing structural properties like containment, reference, cardinality, etc., among the participating components. To mitigate this drawback, LSCs need to be supported by additional structural modeling language that can capture some of these richer constructs. Clafer, on the other hand, subsumes the capabilities of LSCs through its ability to model architecture and its support for rich structural properties. For instance, in our case study, the fact that the `express up` button is present only in the case of `express` feature, can be easily stated in Clafer using a single line of expression. Whereas, it would be impossible to capture such a constraint purely in LSCs.

**Behavioural variability (RQ1)**   LSCs, like Clafer, can capture negative scenarios and variability in behaviour, albeit through additional language constructs. For example, negative scenarios can be captured with *forbid* annotation in a chart [52]. Variability of behaviour can be realized via variation points on transitions, coded as a condition inside an *alternative* box. Unfortunately, since LSCs do not have first class support for features, such conditions might have to rely on a complex reflection mechanism to check for the presence of a certain feature. Expressing alphabet variability in LSCs is non-trivial. One can, however, use *monitors* as a trigger for a certain chart execution to capture variability of events. A more general variability appears impossibly to express: a specification of the form that a sequence of behaviour is present only when a certain feature (like `express` in our case) is present, becomes difficult to capture purely in LSCs.

**Tool support for language users (RQ2)**   LSCs are primarily a graphical language, with textual support for describing use-case scenarios. This may contribute to lower cognitive load on designers using the language, especially in early stages of learning it. Although there exists Java API support for creation of LSC models, the benefit of reducing cognitive load on designers is lost if not relying on the graphical interface of the PlayGo tool. LSCs support analysis which involves invoking the PlayGo execution engine. This generate traces of valid execution scenarios in linear time. The PlayGo tool is feature rich with a built in natural language interpreter, code generation, and trace analysis. In contrast, the current behavioural Clafer tooling was designed to offer trace generation and consistency checking; however, it seems that similar tools could be developed for Clafer, given that LSCs are essentially a semantic subset of Clafer.

**Support for language implementers (RQ3)**   Although the language specification is moderate (48 pages), developing tools for LSCs involves significant cost due to the graphical syntax. Input models would have to be created for UML model profile to be compiled by the S2A compiler [37]. Creating tools for Behavioural Clafer involves less overhead. Generic editors and version control tools can be used. One needs to develop a compiler to a solver. The core language is small and extensions can be realized by simple syntactic sugar definitions.

### 7.2 Systems Modeling Language (SysML)

The *Systems Modeling Language*, or SysML for short, extends the UML specification to enable system engineers to model structure and behaviour of systems in general, not necessarily restricted to just software components. System components like *ports*, *connectors* and *switches* are first class entities along the software components in SysML [31].

**Structural variability (RQ1)**   SysML includes variability modeling mechanisms, although their use is not simple. Admittedly, the complexity is partly caused by the overall richness of the language, not only by the said mechanisms directly. We illustrate this with an example.

Accordingly, figure 7a shows a SysML model of component decomposition for the power window case study. Its primary purpose is to show the system decomposition into the controller, the chimer and the switch. The diagram also expresses some variability. Observe that `Controller` is further decomposed into an express controller `ExpController` and a `BasicController`, which are meant to be exclusive choices, like in our Clafer model. Both containments have cardinality 0..1, which makes them optional (similarly to how we did it in the Clafer models earlier). This second level of decomposition could, potentially more naturally, be modeled using generalization. In order to illustrate both patterns, we show the generalization solution in the implementation of variability of the switch: `ExpressSwitch` and `BasicSwitch` both specialize the switch block in the same diagram.

Now we use *bounded references* (introduced in SysML 1.4 specification) to capture the constraints and specify valid variants that can be created out of the model. We first map each component with variability to a dedicated bounded reference (figure 7b). Finally, the system with generalized bounded reference is specialized into specific variants as shown in figure 7c. In the figure, we see how we can redefine the generalized bounded references to specify two valid variants, one with express controller `ExpController` and no `Chimer`, and another with `BasicController` and `Chimer`. Notice the cardinality overrides to 0 and 1, respectively, in each of the variant boxes.

A notable advantage of SysML though is that designers can create system models that reflect low level system design or hardware architecture more accurately. This decomposition from a higher abstraction to a lower one is usually achieved by allocating components like in figure 7a to separate *block definition diagrams*. Components in such block diagrams are further allocated to *internal block diagrams* and separate behavioural diagrams like *activity diagrams, state charts* etc. Furthermore, figure 7

**Clafer: Lightweight Modeling of Structure, Behaviour, and Variability**

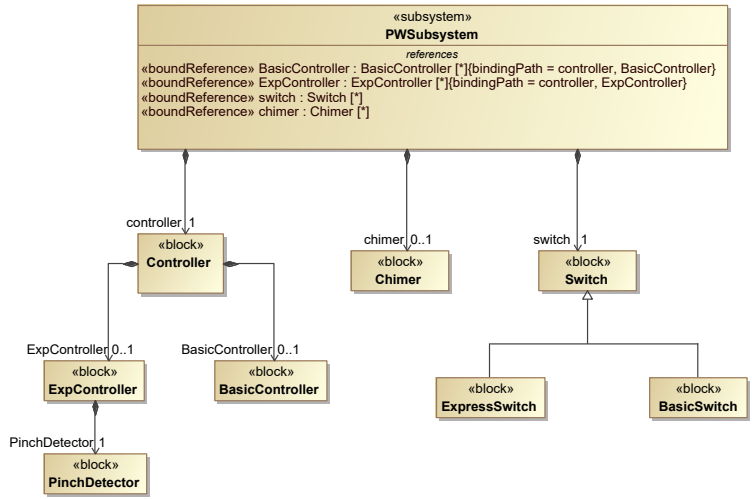

**(a)** SysML: Decomposition hierarchy of the power window subsystem into its individual components along with the bound references to corresponding variants.

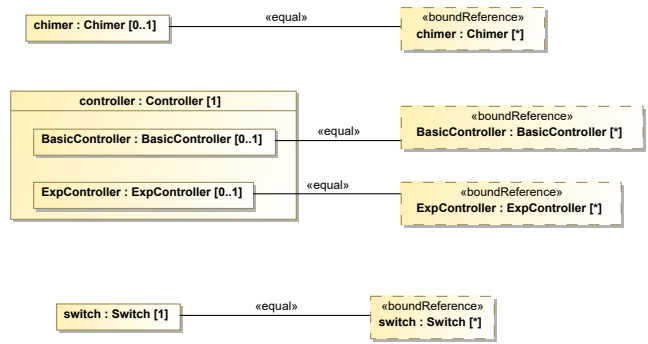

**(b)** SysML: Mapping of each component with variability (Chimer, BasicController, ExpController) to dedicated bound references.

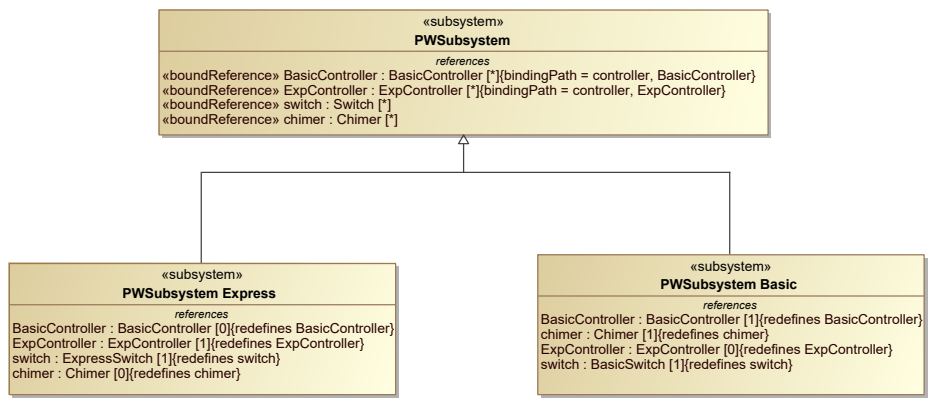

**(c)** SysML: Creation of concrete valid variants (ExpController and BasicController) by assigning appropriate bound reference relevant to individual variants.

**Figure 7** SysML structural decomposition

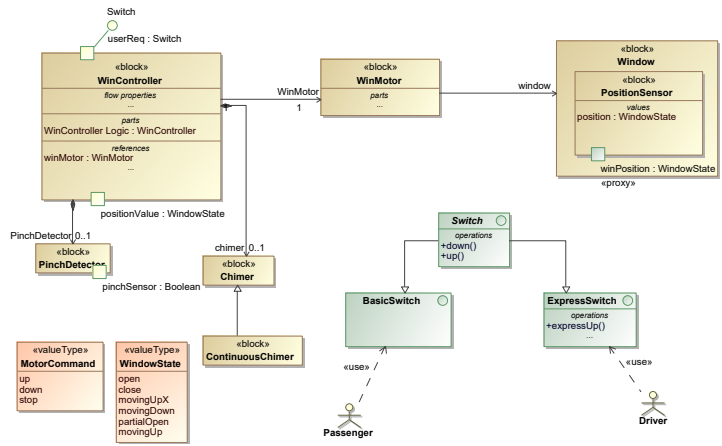

**(a)** SysML: Complete block definition diagram (BDD) of the power window subsystem showing all the relevant blocks, references, and value types.

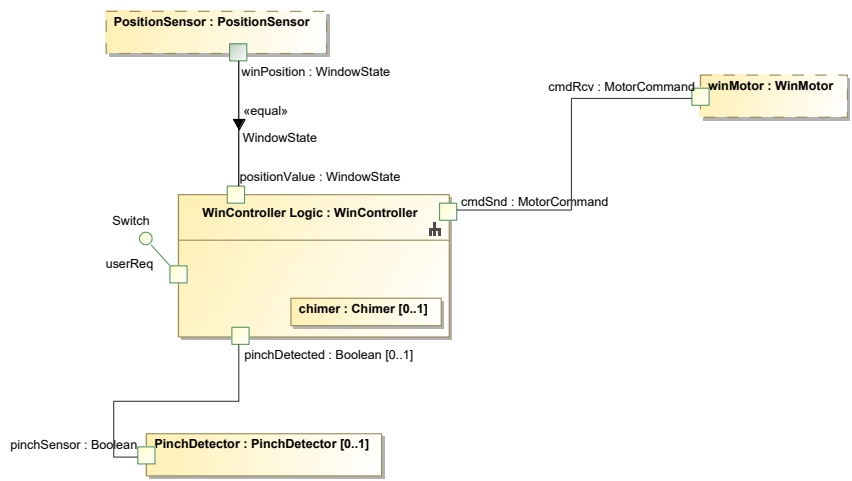

**(b)** SysML: Internal block diagram (IBD) that shows the internal structure of WinController block defined in the BDD.

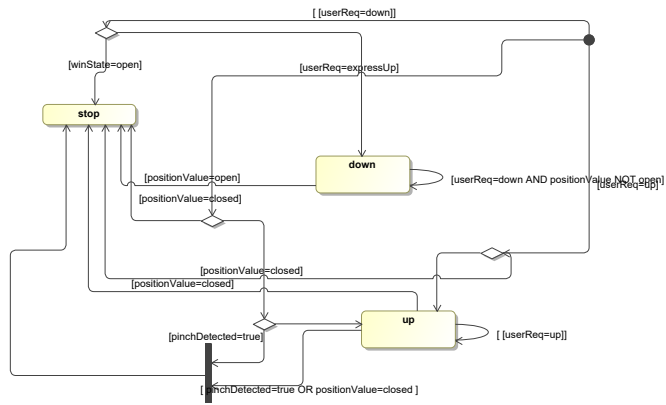

**(c)** SysML: State machine diagram that describes the behavior of the WinController IBD.

**Figure 8** SysML behavioural modeling

shows an example of the block definition diagram of the power window subsystem, the internal block diagram of the controller and the corresponding state machine. Although such diagrams capture variability through different variability patterns [32], the compositional nature of figure 7a is lost in figure 7. Inevitably, this results in organizations having to maintain separate models with complex relationship among each other, often leading to inconsistencies. Speaking to our industrial collaborators, this overhead is one of the major hindrance towards wider adoption of SysML.

Compared to Clafer, the steps involved in modeling structure with variability and in configuring variants, generate significantly more cognitive effort for the designer. Whereas in SysML, the designer has to maintain at least six different models (three for configuring valid variants from feature models, one each for block definition diagram, internal block diagram, and state machine), Clafer can express all these in a single model purely through the language constructs. Also, modeling variability in behaviour and tracing them to structure is much more complex in SysML than in Clafer.

**Behavioural variability (RQ1)**   The contrast between SysML and Clafer is further highlighted while modeling variability in behaviour. As shown in section 4, Clafer enables the designer to express the different patterns of variability compactly. In SysML, there is no direct way to capture these patterns of behavioural variability. Alphabet variability, for instance, needs to be captured in the following way. Since a valid variant of figure 7c restricts the `Switch` type to `BasicSwitch`, which does not have the option for `expressUp` command (figure 8), any state transition diagram allocated from a `BasicSwitch` would not have `expressUp` in the alphabet. Similarly, even for transition variability, one needs to go through the complex layers of allocations, to capture the constraints in behaviour. For a designer who is well versed with the SysML specifications and tooling, this might be intuitive. However, for novice or less experienced designers, this relationship becomes overwhelming.

**Tool support for language users (RQ2)**   Currently, there are close to dozen SysML editors in the market, most of which are commercial.[18] SysML helps the designers using graphical syntax and offering directly low level abstractions that don't need to be created by the modeller (for instance in Clafer, one needs to create the concept of a port). However, as discussed earlier, this advantage is offset to a degree by the need to maintain multiple disparate diagrams, and by the size of the language, incurring a high entry barrier.

**Support for language implementers (RQ3)**   The ability to model low-level abstraction in SysML comes at a cost of verbosity in the language specification (250 pages). This results in significant investment in developing tools and editors. Probably, this is the main reason why mostly commercial tools exist (no lightweight open source tools).

---

[18] http://sysml.tools/, last accessed on 2018-07-19.

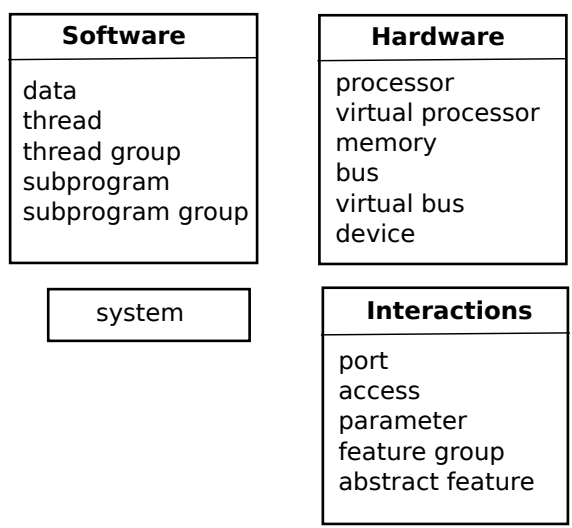

**Figure 9** Primary AADL components

```
process WinController
 features
   current : in data port
   position : in data port
     .
     .
 end WinController;
            (a)

process implementation WinController.basic
subcomponents
   thr_control : thread WinController_thr.generic
end WinController.basic;

process implementation WinController.express
subcomponents
   thr_control : thread WinController_thr.express
end WinController.express;
            (b)
```

**Figure 10** AADL variability through implementation

### 7.3 Architecture Analysis and Design Language (AADL)

The *Architecture Analysis and Design Language* (AADL) [27] is a textual architecture description language, developed by SAE International, initially targeting the avionics domain, but now aimed for embedded systems in general. AADL supports design and analysis of hardware, software and their interface. Moreover, figure 9 summarizes major AADL components used for modeling structure and behaviour. The various categories allow creating both low level architectural models and abstract models. The AADL models are used for diverse analyses such as timing, safety, etc. However, like other architecture description languages, due to its detailed specification, AADL comes with a steep learning curve for designers [50].

```
thread WinController_thr
features
 current : in data port;
 winReq : in event data port;
 object : in data port;
 winCmd : out data port;
 position : in data port;
flows
 .
 .;
end WinController_thr;

thread implementation WinController_thr.generic
 annex behavior_specification {**
  states
   halt : initial complete state;
   init, cmd_up, cmd_down : state;
  transitions
   halt -[on dispatch]-> init;
   init -[ position!=1 and winReq=1 ]-> cmd_up {winCmd:="up"};
   init -[ position!=0 and winReq=0 ]-> cmd_down {winCmd:="down"};
   cmd_up -[winReq=2]-> halt {winCmd:="stop"};
   cmd_down -[winReq=2]-> halt {winCmd:="stop"};
 **};

end WinController_thr.generic;

thread implementation WinController_thr.express
 annex behavior_specification {**
  states
   halt : initial complete state;
   init, cmd_up, cmd_exp_up, cmd_down : state;
  transitions
   halt -[on dispatch]-> init;
   init -[ position!=1 and winReq=1 ]-> cmd_up {winCmd:="up"};
   init -[ position!=0 and winReq=0 ]-> cmd_down {winCmd:="down"};
   cmd_up -[winReq=2]-> halt {winCmd:="stop"};
   cmd_exp_up -[winReq=2 or object=1]-> halt {winCmd:="stop"};
   init -[ position!=1 and winReq=3 ]-> cmd_exp_up {winCmd:="up"};
 **};
end WinController_thr.express;
```

**Figure 11** AADL behavioural modeling

**Structural variability (RQ1)** AADL supports variability modeling by separating component *types* and component *implementations*. Component types are definitions listing the component's elements, interfaces and externally observable elements. Component implementations, on the other hand, specify the internal structure and execution profile. Accordingly, there are two ways of implementing variability in AADL: by redefining types and by providing alternative implementations.

Variability by type redefinition exploits the `extends` and `refined to` constructs of the language. This is similar in spirit to the common approach of variability modeling through specialization, where new component types are defined as substitutable for a more general type. Unfortunately, compared to Clafer and SysML, the absence of group cardinality support for subcomponents, makes this kind of variability modeling restricted in AADL. Shiraishi shows how this pattern can be used to model variability in automotive cruise control system [67].

Variability by alternative implementations is similar in nature to type variability but happens at instance level modeling. Different variants can be composed to have different implementations, varying in their internal structure. Also, figure 10 shows an example of two different implementations of the `WinController` process, one for the `basic` and one for the `express` controller.

Both patterns express variability solely on the base architecture model. AADL does not support modeling or integration with external variability models like feature models or OVM. This drawback makes the process of concrete product derivation non-trivial, and there has been dedicated research effort to integrate the two [33, 51]. In contrast, in Clafer this integration comes for free, given that feature models are naturally implemented using clafers and constraints.

**Behavioural variability (RQ1)**    The core language specification of AADL does not support variability in behaviour. This may be achieved using the behaviour annex [24]. AADL behaviour specification is based on a state transition system, and can be included in a component implementation.

In the power window example, the execution behaviour is modeled as a part of thread implementation of the `WinController` process (figure 11). The figure shows two thread implementations, one each for express and basic controller. In the express implementation, there are additional transitions for conditions when the user presses the `express up` button (`winReq=3`) and a pinch object is detected (`object=1`). There is also an additional state (`cmd_exp_up`), which results from the additional transition.

AADL supports modeling behavioural variability only through alternate implementations, where variability is modeled primarily at alphabet level (*alphabet* variability). AADL's inability to provide out-of-the-box integration of its behaviour models to external variability models, prevents the designer to implement fine grained variability, for example using the transition variability pattern.

**Tool support for language users (RQ2)**    There exist several tool-chains supporting AADL, most of which are based on OSATE 2 open source platform. The platform provides an IDE for development and analysis of AADL models. In the basic OSATE 2 IDE, AADL models are created in textual format, with a support for graphical representation of the architecture model. There also exist tools developed based on a graphical platform.[19] Most tools cater specifically to the domain of real-time embedded systems, providing timing and safety analyses. Clafer supports no real-time analyses so far, but safety invariants can be expressed as assertions.

---

[19] https://wiki.sei.cmu.edu/aadl/index.php/AADL_tools, last accessed on 2018-07-19.

**Clafer: Lightweight Modeling of Structure, Behaviour, and Variability**

**Support for language implementers (RQ3)**   The language specification [27] is extensive: 145 pages *excluding* annexes such as behaviour annex, data-modeling annex, etc. Many extensive languages are included in the annexes, most tied to a specific domain (real-time embedded systems). For example, the specification of BLESS, a behavioural language based on AADL, is 200 page itself.[20] The proliferation of dialects, and their complexity clearly makes implementing AADL difficult, although this is partially offset by existence of basic platforms, which help in scaffolding more complex tools.

### 7.4  Temporal OCL

*Temporal OCL* is an extension of the *Object Constraint Language* allowing specification of temporal constraints on object models. In principle one could express temporal constraints directly in OCL, but doing this in first order logic is cumbersome, and moreover OCL limits the use of quantification over temporal domain (allowing only expressing universal invariants, or pre-post conditions over single transition step). This motivates development of separate timing constructs in the constraint language. Several variations of such temporal extensions to OCL exist. We consider the work of Kanso and Taha, as one of the more recent attempts [44]. Their tools[21] supported modeling the case study for evaluation. We remodelled the example using Ecore and wrote the constraints in OCL.

In many respects, the semantics of Temporal OCL is similar to behavioural Clafer. Both languages model structures as MOF-like diagrams (roughly for Clafer) and add behaviour as constraints on temporal changes to structures. Also both languages are based on Dwyer's property patterns, in order to simplify the specification of typical

---

[20] http://www.santoslab.org/pub/bless/docs/BLESS_Language_Reference_Manual.pdf, last accessed on 2018-07-19.
[21] http://wdi.supelec.fr/software/TemporalOCL/, last accessed on 2018-07-19.

■ **Table 1**  Comparison of modeling languages based on variability

|  | Structural variability (RQ1) | Behavioural variability (RQ1) | Specification size (RQ3) (pages) |
| --- | --- | --- | --- |
| Behavioural Clafer | Full support integrating feature-models, architecture and behaviour | Alphabet and transition variability | <20 |
| Live Sequence Charts (LSC) | No support | Expressing alphabet and transition variability is non-trivial | 48 |
| SysML | Partial support; variability cross-cuts several model views | Expressing alphabet and transition variability is non-trivial | 250 |
| AADL | Partial support; through type redefinition and multiple implementations | Only alphabet variability | >145 |
| Temporal OCL | No intrinsic support | Alphabet and transition variability | <20 |

```
context controller
 temp trans_var : isCalled(controller_exp_up())
      preceding directly becomesTrue(self.psensor.state
      =win_state.exp_up)
       when self.express_feature->size()=1
                1. Temporal OCL

[ basic -[express && req=expressUp]-> movingUpX ]
                2. Behavioral Clafer
```
**(a)** Transition variability

```
context controller
 temp alph_var : never becomesTrue(self.psensor.state
      = win_state.exp_up)
       when self.express_feature->size()=0
                1. Temporal OCL

[ some WinFeatures.express <=> some expressUp ]
                2. Behavioral Clafer
```
**(b)** Alphabet variability

■ **Figure 12** Modeling behavioural variability patterns in Temporal OCL and Clafer

temporal constraints. We modeled features using containment hierarchy similarly to what was presented earlier in this paper and in our earlier work [4].

**Structural variability (RQ1)**  The objective of Temporal OCL extensions is solely to provide temporal capabilities to OCL. For structural variability one can use regular OCL over class diagrams in similar style to Clafer, perhaps at the cost of more verbose notation. This topic has been extensively discussed in our earlier work [4]. Jackson's book [42] offers an extensive comparison of Alloy's constraint language and OCL, which largely translates to Clafer, which uses the same constraint language (but a more compact structural language than Alloy). Notable advantage of Clafer is group cardinality, which has to be encoded in OCL and Alloy.

**Behavioural variability (RQ1)**  Temporal OCL is the closest to Clafer out of the discussed languages, as far as modeling behavioural variability is concerned. As shown in figure 12, Temporal OCL can capture both *alphabet* and *transition* variability patterns. Since we model `express` feature as containment with the `controller` class, in figure 12, the specification does not need to depend on reflection mechanisms. However the constraints appear much more complex than in Clafer (see figure 12). Temporal OCL's specification borrows heavily from OCL. Due to this, designers need to master OCL specifications first, which by itself has a steep learning curve [14]. Clafer, on the other hand, provides syntactic sugar and defaults that allow even novice designers to model complex scenarios. Another drawback of Temporal OCL is the separation of structure from specification. This often leads to a dependence on heavy tooling support; a tool for integrating external variability models with structure, another one for integration of structure with specification. Clafer minimizes this dependency by being tool independent. The models can be created and edited in any textual editor, and it provides seamless integration of feature models, structure and behavioural variability solely through the language constructs.

**Tool support for language users (RQ2)** Temporal OCL provides an Eclipse based TOCL editor, where designers can express Temporal OCL constraints over an Ecore model. Analysis capability of the tool is limited, where execution of the specification constraints generates an equivalent regular expression. Clafer, being based on a more carefully delimited language, benefits from tools based on SAT solving, supporting consistency checking and snapshot generation for static subset. None of these are presently supported for Temporal OCL.

**Tool support for language implementers (RQ3)** The size of language specification is minimal, compared to the languages discussed above. This makes the learning curve, at least in terms of language specification, similar to that of Clafer. Temporal OCL needs to be supported by a structural modeling notation (class diagrams) which complicates the implementation, but can also be a benefit if the implementer is well versed in available frameworks providing such (for instance Eclipse).

## 7.5 Summary

A short summary of the results of this case study is presented in table 1. We see that Clafer presents itself as the language that is both very small and fairly expressive, albeit the expressiveness is achieved by patterns, and not directly. Thus, regarding RQ1, Clafer tends to offer more than competition, but at the cost of learning extra language knowledge—the modeling style (patterns). Terseness of Clafer models stands out in comparison with other languages. This also means that embedding of structure and variability with modeling patterns does not cost syntactically. Regarding RQ2 we observe that Clafer shares most strengths of other textual languages and languages with constraint based semantics. At the same time, it also suffers from the cost of the semantics, given that execution requires solving difficult algorithmic problems (for instance adding real-time support is predictably difficult). Finally, regarding RQ3, Clafer presents a very concise definition, and is likely easier to be implemented than many other languages.[22]

In section 8, we reflect on the design decisions in Clafer, emphasizing how do they contribute to this result. We also point out shortcomings and difficulties.

---

[22] For Temporal OCL, we excluded the size of the core OCL specification, which is over 250 pages (https://www.omg.org/spec/OCL/2.4/PDF, last accessed on 2018-07-19.)

**7.6 Limitations and Validity**

The method followed in this experiment combined *design* of models (artifacts) within a *case study* (Power Window) and a *qualitative analysis* of the created and available *artifacts*. We discuss the main issues and limitations that this method and its execution incur, in order to estimate the trustworthiness of results.

**Repeatability**   We have reported (above) the languages, the research questions and the comparison dimensions used. The models framing the discussion of RQ1 were designed in a two step improvement process: *create-critique-improve*, which increases a chance that idiomatic use of languages has been achieved. We have stored and published all the models.[23] We have strived to be unbiased in analysis discussions, and we reasonably believe that other researchers departing from the same reference models would reach similar conclusions. However we do not claim that the analysis would always give the same result (as is the case with almost any data analysis process, and especially a qualitative one).

**Generalizability**   This evaluation section should be read as a voice in the discussion on design of modeling languages. Obviously, we worked only with a single model, and with only five languages (including Clafer). The conclusions taken can only be directly traced to this context. Comparison with other languages and in other use cases could give different results. We believe though that given the relatively diverse spread of the languages considered, and the richness of the case study (showing various aspects of modeling) the observations are useful. They also convince us that Clafer lies in a relevant design point in the space of languages.

**Bias**   The design of Clafer is independent of the created case study. The language had been designed using smaller toy examples, and the case study was only introduced when the key principles where stable. Thus Clafer design is not overfit for the power window case. The modeller working in the four baseline languages was not involved in creating the Power Windows case in Clafer, and was not not part of the team designing Clafer. This reduces a chance of bias towards Clafer in creating the four cases, but we admit it does not entirely remove it. The language experts used to critique the created models were recruited both from among the coauthors and external lab members. This obviously introduces a possibility of contamination with bias, that is overall very hard to avoid in a qualitative comparison of languages.

**Metric representativeness**   We realize that the size of the documentation is a rather poor metric for language complexity. Some pages may be less or more useful, and the density of information varies a lot. Also, the size of the documentation is polluted by the age of the language (short living languages are more likely to have shorter documentations due to shorter accumulation time). However language complexity is a very fragile concept, and this was a metric that is relatively stable and unbiased, among the scarcely few that exist. Even needless pages introduce a burden on users and implementers, who need to find information. Thus we decided to report it nevertheless. We do not derive any conclusions from precise values of the metric, but focus on the order of magnitude (tens vs hundreds of pages).

---

[23] https://uwaterloo.ca/waterloo-intelligent-systems-engineering-lab/publications/clafer-lightweight-modeling-structure-and-behaviour, last accessed on 2018-07-19.

Furthermore, the cost of language implementation should, in general, be correlated with demand for its features and with benefits it offers, so we advice the reader to not jump to fast conclusions based on our comparison.

## 8 Discussion

We have experimented with Clafer, creating more models. Besides the power window model, the paper material[24] contains other cases available for the research community:

- *TrafficLight.* A simple state machine model for traffic light transitions.
- *WebSocketProtocol.* A model of the IETF RFC 6455 protocol [61].
- *TLS Handshake* from the IETF encrypted communication protocol (TLS/SSL) [60].

We summarize the observations made in the process of designing the behavioural extension of Clafer and using it for modeling. We list its main shortcomings and envisioned future work. Methodologically, the observations presented in this section differ from those of section 7 in that they are all based on authors own experience and reflection in the design process and early use; in contrast, to the observations in the previous section that were collected through modeling in several languages and comparing the models and language documentations. In that sense, the list below is much less systematic, and more bias prone. Yet, we believe still worth bringing for the language engineering audience.

**Lesson 1: Modeling behaviour does not require a large language extension**   Clafer, a rather small structural modeling language to begin with, required only two additions to the core syntax: the LTL operators next and until. A larger change was required in the semantics, which was lifted from sets of structures to sets of traces of structures. This is in stark contrast to, for instance UML, where large sub-languages have been added to handle behavioural modeling.

**Lesson 2: Nesting and behaviour interact efficiently**   Mixing the composition hierarchy and behavioural modeling elements simplifies constraints significantly. In our experience, most of simple behavioural properties (as met in studied cases) reduce to static invariants nested in a suitable context, which itself is dynamically activated. Besides this we also use transition constraints a lot, but we only rarely see the need to use LTL properties or advanced property patterns directly.

---

[24] http://www.clafer.org/2018/05/behavioral-model-set.html, last accessed on 2018-07-19.

**Lesson 3: Variability modeling can be efficient without first-class support in the language**
Clafer's ability to express variability in structures naturally extends to behaviour. We demonstrated several patterns for modeling variability in behaviour: super-imposed variability, alphabet variability, transition variability, behaviour refinement via inheritance, and use of the strategy pattern. We used types, inheritance, references and the ability to under-constrain models, and no explicit notion of features, variation points or mapping links. We used LTL (syntactic sugar for `final`) to model static binding time, frequent in variability modeling. In fact, we are not aware of any other modeling language, in which all these patterns could have been demonstrated with comparable ease.

**Lesson 4: Looseness in Clafer models has many applications**   Under-constraining of structure and behaviour in models does allow not only creating variation points controlled using constraints, but introduces tolerance for uncertainty, and underspecification [5]. Thus it can also support other use cases.

**Lesson 5: Integrated notation allows obtaining compact models**   Different aspects of a system can be represented in a single place. The running example, which fits on a single page, takes at least two diagrams in Simulink or UML. Scenarios and properties need to be specified in a yet separate place. A language like CVL or OVM, would add additional diagrams (feature models, feature mappings) just to maintain variability on top of the UML or Simulink model. Furthermore, an integrated notation does not enforce compartmentalization of the modeling process. The modeller can freely mix modeling structure, behaviour, features and variation points, and indeed we did so in the running example. This is not impossible with multiple notations, but compartmentalization often happens unconsciously as switching between notations creates a cognitive burden. Importantly, the opposite is also true: creating large amalgamated models encompassing multiple aspects also increases this burden. Yet, a language like Clafer allows the user to choose whether to model variability within structure/behaviour or whether to separate it to a feature model and link with structure/behaviour using constraints.

**Lesson 6: Integrated notation allows performing integrated analysis**   We demonstrated that variability can be added to different types of models. Similarly, behaviour can be added to traditionally static models, such as, feature and architectural models. The freedom of expressing both invariants and behaviours over many types of structures opens new possibilities, modeling feature model configuration workflows, such as, staged configuration, or expressing dynamic aspects of architectures.

**Lesson 7: Combining both bottom-up and top-down modeling opens up new possibilities**
Traditionally, the behaviour of a system is expressed imperatively in an automata-like notation, while the requirements to be checked are specified declaratively as properties. Clafer supports both styles of modeling; however, both can be used for either purpose. That is, the system behaviour can be specified declaratively using properties and patterns, while properties to be checked can be specified more imperatively using scenarios. This flexibility opens up possibilities for developing new modeling methodologies such as example-driven modeling [6] and using appropriate means to express what is required. For example, *state properties* are often easier to write as properties than as transitions, while, other properties maybe easier to write as scenarios.

**Shortcomings**   We realize that a very tight language has its limitations. It might not always be easy to see for what purpose a construct is used. For instance in our models multiplicity `0..1` has been used both to express optionality (`express`) and that a state is not fixed and can change (`movingUp`). We mitigated this problem by using types (such as `Feature` or `State`) to capture intentions explicitly. Naming conventions can also be used as a lightweight mechanism.

Clafer cannot capture all styles of variability modeling. It is not very well suited for modularization of differences between variants in style of delta-modeling [64] or FORML [66]. The main obstacle is that a clafer must be made under-constrained in order to make it variable. In the differential modeling style, a model element can be fixed in the model, but still changed by an external difference module, which has the ability to "edit" base models, which is not possible in Clafer.

The constraint-based semantics is rich enough to express many styles of models, but it can lead to implicit information in models. For instance, links between model elements and features are not explicit, and identifying them requires using analysis tools. Similarly behaviour specified using properties might be quite indirect, especially when not using the pattern for states and transitions. In Clafer, system simulation requires solving consistency of LTL formulae. However, the recent advances in SAT and SMT solving as well as in bounded model checking bring us much closer to making such languages usable.

**Future work**   First, we would like to explore different ways of using Clafer models with behaviour. A behavioral model could be interpreted as a static model by ignoring all temporal constraints and current instance generation and consistency checking capability of Clafer backend reasoners could be used to explore the snapshot space of the model. Next, a behavioral model could be used for trace generation and model checking. Lastly, a behavioral model could be used for generating a property checker for run-time monitoring.

In a direct continuation of this work, we would like to explore possibilities of capturing more modeling styles in Clafer. We are interested in full-fledged sequence diagrams (as opposed to simple scenarios), and in manipulating the alphabet of a system more subtly than just deactivating inputs [48]. Building Petri-net like models is another avenue to explore: state multiplicity can be relaxed from `0..1` to higher bounds allowing for multiple instances of a single state at a time, corresponding roughly to the number of tokens present in a state.

We would also like to explore designing modeling methodologies that exploit Clafer's flexibility, especially deciding what aspects are better to explain using transition models, and what aspects are easier to do using properties; how models should be debugged; and how scenarios and assertions should be selected to achieve a good quality check. Finally, we need a number of visualization and analysis tools, including visualization of models as automata, generation of transition systems, efficient dedicated model checkers and simulators for Clafer.

We do not claim that Clafer invalidates all multi-modeling approaches that use specialized notations for different aspects of the system. Clafer is a voice in discussion where the boundaries should be placed between languages and what should be part of the language and what part of the model. What kind of syntactic and semantic elements can help making languages smaller but models bigger? This is related to the old discussion between virtues of internal and external DSLs. It would be extremely interesting to advance this discussion using a controlled experiment, for instance Comparing Clafer (uninotational) with a multi-language approach (say CVL combined with UML Class Diagrams, State Diagrams and Sequence Diagrams). Having participated both in Clafer and CVL design, we are quite prejudiced about the potential result. However, such an experiment would very likely uncover a number of interesting areas where Clafer is actually weak, and where the multi-notation approach shines. Perhaps, it would even be able to identify what aspects are better handled in the language, and what in the model.

Furthermore, a controlled experiment would help to identify issues regarding adequateness, ease of learning, and ease of use for Clafer. No formal experiment of this kind has been executed. However, the structural part of Clafer has been used in teaching both University of Waterloo, at IT University of Copenhagen, and at DSM-TP modeling summer school (four times). We estimate that more than 200 students have been exposed to Clafer and we have been grading the models produced by them in our interactive wiki. Although informal, this experience indicates that both undergraduate and graduate students assimilate Clafer fairly easily, and easily gain understanding of modeling (structural) variability through patterns of structures and constraints. We also interacted with two European companies that experimented with Clafer for their use cases independently (not our collaborators), which indicates that the language was attractive at first sight. So far, behavioral part of Clafer has only been taught in passive manner (lectures) and not in an active way (interactive tutorials) as for the structural part. This means that the existing experience regarding behavioral modeling in Clafer is limited and further experimentation would be very valuable.

## 9  Related Work

In the interest of conciseness, we limit the discussion of the related works to papers that have not been extensively discussed before.

**Modeling and Semantics at Large**   The field of model-driven development produced a number of valuable tools and formalisms for modeling structure and behaviour of software systems. For instance, the class diagram formalism [49] popularized by UML has gained considerable industrial acceptance for *structural* modeling [41]. However modeling behaviour in UML requires switching to other notations, with all the cognitive cost that it takes, and all the unclarity introduced by mixing multiple languages. Another example, the PROMELA language of the SPIN model checker [40] has been used in many *behavioural* modeling projects. PROMELA presents excellent facilities for modeling dynamics, however its support for modeling data or structure barely exceeds simple data types. More recently, Alloy [42] introduced modeling of *both* behaviour and rich structures, by allowing to encode behaviour using structural notation and transitive closure (while expressive this is quite unlike the popular intuitive notations known from automata theory or temporal logics).

There is a large body of works on extending the Object Constraint Language[25] (OCL) with behavioural operators. We discuss selected examples here [15, 29, 44, 72]. Consult Kanso and Taha [44] for a more thorough recent comparison. Temporal extensions of OCL available in the literature have very different goals from Clafer. They introduce a large number of first class concepts to capture temporal aspects, including timestamps, clocks, timers, deadlines, and signal handling. Most also employ some special syntax for accessing past values of variables. Our goal is to provide a minimal behavioural extension of the language that enables to model most common use cases. We only add new constructs when indispensable (for instance accessing past variables in Clafer is done using a standard let binding, without introducing special temporal variable accesses). Clafer remains small, to ease adoption, to be useful in teaching, to enable building maintainable tools, and to facilitate formal studies of models mixing behaviour and structure.

Trace semantics, similar to Clafer's, is seen in some of these works [15, 44, 72]. Flake and Mueller [29] use Clocked CTL, which requires a tree-based semantics. However, none of these works uses temporal logics to *model* behaviours. All temporal extensions of OCL known to us are used to write property *specifications* to be checked against statechart models. For this reason the syntax of these extensions is not developed towards direct behaviour modeling, and it is essentially impossible to apply the modeling patterns like in our work (it is theoretically possible, but the obtained models will not reflect the behaviours directly, but will use a hard to understand encoding). Besides, none of the mentioned work considers using OCL with behaviour to model variability.

Interestingly Cengarle and Knapp [15] include handling of past operators of LTL, which we consider useful in behavioural modeling (even if not increasing expressiveness). Past operators would be a beneficial extension for Clafer, but primarily for writing properties (assertions) not for modeling behaviour. Flake and Mueller [29] only apply temporal specifications to configurations of statecharts states, so they cannot reason about evolutions of object-oriented structures.

Two frameworks use Alloy to model behaviour by exploiting transitive closure over structural relations. Cunha [17] shows how to achieve bounded model checking of LTL in Alloy using a local state idiom. Vakili and Day [68] present an Alloy library providing universal and existential bounded model checking of CTL* with fairness constraints. They use the global state idiom and assume a direct representation of a one step next-state relation, which makes the set up a bit hard to use for Clafer, in which behaviour emerges from composition of constraints over traces. Both proposals are remaining within Alloy, which is a serious advantage for Alloy users, but it makes it impossible to stylize models into domain-specific sub-languages like we do (for example to express feature models, hierarchical state machines or scenarios), due to lack of nesting, intelligent name resolution, and syntax for modeling behaviours.

---

[25] http://www.omg.org/spec/OCL/, last accessed on 2018-07-19

DynAlloy [30] is a first-class extension of Alloy with dynamic semantics based on traces. It allows specifying pre- and post-conditions for traces of actions (partial correctness assertions for traces). Axiomatic reasoning in this style is not very well suited to modeling automata-like control to support use cases in modeling embedded systems and business processes, in which we are interested.

**Variability Modeling** As mentioned, there are two main patterns for design of variability modeling languages: an *amalgamated* (or *integrated*) approach that extends an existing language with variability mechanisms, and *separate* (or *orthogonal*) approach that keeps all the variability aspects outside the modeling language. This also usually means that modeling of individual features may be made modular. These two patterns are mixed to an extent in all designs.

FORML [66], Delta modeling [64], Orthogonal Variability Models [58], and the Common Variability Language [39], are the main examples of this modular approach. These are inspired to a large extent by aspect-oriented programming. Indeed, aspect-oriented programming has been used directly in variability modeling and implementation, for example refer to [3, 34]. As we already discussed, they all differ from Clafer, in being independent variability modeling languages, imposed on top of an existing general modeling language, while Clafer is an integrated language, with no first-class support for variability, but where variability can be represented using design patterns.

With the exception of FORML, the above languages do not support behavior directly (all the above languages support behavior indirectly, by manipulating syntax trees of behavioral models). A number of works have been devoted to extensions of semantic models and behavioral modeling languages with variability. The earliest examples include extensions of transition systems for variability modeling [16, 28, 47, 48]. The difference between Clafer and the existing work lies in combination of structural and behavioral modeling in a single language. This is a rather rich area of research. We direct the reader to a very good recent survey on this topic, listing over 40 papers [7].

The structural version of Clafer has been presented before [4]. The present paper contributes a very non-trivial extension with the temporal dimension. Modeling behavior in the original structural Clafer was (almost) impossible—while one can always encode behavior in structural properties, in this case the encoding turned out extremely heavy weight. The behavioral extension required integration into the Clafer semantics the temporal aspects following the foundational work of Abadi and Manna [1] and into the Clafer compiler the bounded model checking machinery [12, 17]. Conceptually, section 3 in the present paper overlaps with the original work of Bąk, Diskin, Antkiewicz, Czarnecki, and Wąsowski [4]. However, the addition of behavior required some interventions in the static part of the language as well (clafer modifiers `initial` and `finally`). Thus we decided to present the static part of Clafer as well, making the present paper self-contained and up-to-date. We also created an entirely new presentation of this material, emphasizing the use of patterns, consistently with section 4. This means that the contribution of section 3 lies in modeling methodology rather than in language design. Furthermore, the original categorical semantics [4] was rather cumbersome to extend towards behaviour (more work is required in that direction), thus we decided to use Abadi's scheme for mixing linear temporal logics with first order logics. As a result the semantics presented in section 5 is entirely new with no conceptual overlap with earlier papers. So are the evaluation and discussions sections.

Ross, Murashkin, Liang, Antkiewicz, and Czarnecki [63] present an application of *structural* Clafer for design synthesis. Similarly, Nadi and Krüger report on modeling cryptographic components in structural Clafer [55]. Besides using Clafer, these works do not overlap with the present text.

## 10   Conclusion

We have presented an extension of Clafer with behavioural semantics and demonstrated how this small language captures multiple styles of modeling in a single terse notation (structural, behavioural and variability modeling). To the best of our knowledge, this is the first such language in which demonstration of so many approaches to modeling variability is possible directly. Clafer is also likely the first language which allows *modeling* evolutions of structures (as opposed to only *specification of properties* for evolution of structures), which is a crucial development direction for modeling languages, given the rapid convergence of rich architectures and control in embedded systems.

As we have seen in the numerous examples presented in this paper, despite being quite versatile, Clafer does not seem to suffer a lot from verbosity of models (as would be the case if modeling directly in logics or lambda calculus). Clafer models are meant to be human readable.

Clafer's definition is very concise (if a bit terse), which makes it attractive for teaching and research. It allows to demonstrate many aspects of modeling in one notation to students. Due to a small definition it is also relatively approachable to researchers who would like to work on semantics, languages and tools for models of evolving structures.

**Acknowledgements**   We thank Derek Rayside and Nancy Day for their valuable feedback on the material and earlier versions of the paper. This work was partially supported by The Danish Council for Independent Research under a Sapere Aude project, VARIETE. We thank anonymous reviewers of earlier versions of this work for useful and constructive feedback.

## References

[1] Martin Abadi and Zohar Manna. "Nonclausal Deduction in First-Order Temporal Logic". In: *Journal of the ACM (JACM)* 37.2 (1990), pages 279–317. ISSN: 0004-5411. DOI: 10.1145/77600.77617.

[2] Rajeev Alur and Mihalis Yannakakis. "Model Checking of Hierarchical State Machines". In: *Symposium on the Foundations of Software Engineering*. 1998, pages 175–188. DOI: 10.1145/288195.288305.

[3] Sven Apel, Don S. Batory, Christian Kästner, and Gunter Saake. *Feature-Oriented Software Product Lines - Concepts and Implementation*. Springer, 2013.

[4] Kacper Bąk, Zinovy Diskin, Michał Antkiewicz, Krzysztof Czarnecki, and Andrzej Wąsowski. "Clafer: unifying class and feature modeling". In: *Software & Systems Modeling (SOSYM)* 15.3 (July 2016), pages 811–845. DOI: 10.1007/s10270-014-0441-1.

[5] Kacper Bąk, Zinovy Diskin, Michał Antkiewicz, Krzysztof Czarnecki, and Andrzej Wąsowski. "Partial Instances via Subclassing". In: *International Conference on Software Language Engineering*. Springer, 2013, pages 344–364. DOI: 10.1007/978-3-319-02654-1_19.

[6] Kacper Bąk, Dina Zayan, Krzysztof Czarnecki, Michal Antkiewicz, Zinovy Diskin, Andrzej Wąsowski, and Derek Rayside. "Example-driven modeling: model = abstractions + examples". In: *International Conference on Software Engineering (ICSE 2013)*. IEEE, 2013, pages 1273–1276. DOI: 10.1109/ICSE.2013.6606696.

[7] Harsh Beohar, Mahsa Varshosaz, and Mohammad Reza Mousavi. "Basic behavioral models for software product lines: Expressiveness and testing pre-orders". In: *Systems & Control Letters* 123 (2016), pages 42–60.

[8] Thorsten Berger, Rolf-Helge Pfeiffer, Reinhard Tartler, Steffen Dienst, Krzysztof Czarnecki, Andrzej Wąsowski, and Steven She. "Variability mechanisms in software ecosystems". In: *Information & Software Technology* 56.11 (2014), pages 1520–1535.

[9] Thorsten Berger, Steven She, Rafael Lotufo, Krzysztof Czarnecki, and Andrzej Wąsowski. "Feature-to-Code Mapping in Two Large Product Lines". In: *Software Product Lines: Going Beyond - 14th International Conference*. Edited by Jan Bosch and Jaejoon Lee. Volume 6287. Lecture Notes in Computer Science. Springer, 2010, pages 498–499. ISBN: 978-3-642-15578-9. DOI: 10.1007/978-3-642-15579-6.

[10] Thorsten Berger, Steven She, Rafael Lotufo, Andrzej Wąsowski, and Krzysztof Czarnecki. "A Study of Variability Models and Languages in the Systems Software Domain". In: *IEEE Transactions on Software Engineering (TSE)* 39.12 (2013). DOI: 10.1109/TSE.2013.34.

[11] Thorsten Berger, Stefan Stanciulescu, Ommund Øgård, Øystein Haugen, Bo Larsen, and Andrzej Wąsowski. "To connect or not to connect: experiences from modeling topological variability". In: *18th International Software Product Line Conference, SPLC*. Edited by Stefania Gnesi, Alessandro Fantechi, Patrick Heymans, Julia Rubin, Krzysztof Czarnecki, and Deepak Dhungana. ACM, 2014, pages 330–339. ISBN: 978-1-4503-2740-4. URL: http://dl.acm.org/citation.cfm?id=2648511.

[12] Armin Biere, Alessandro Cimatti, Edmund M. Clarke, and Yunshan Zhu. "Symbolic Model Checking without BDDs". In: *International Conference on Tools and Algorithms for the Construction and Analysis of Systems (TACAS)*. Volume 1579. LNCS. Springer, 1999, pages 193–207.

[13] Moisés Castelo Branco, Yingfei Xiong, Krzysztof Czarnecki, Jochen Malte Küster, and Hagen Völzer. "A case study on consistency management of business and IT process models in banking". In: *Software & Systems Modeling (SOSYM)* 13.3 (2014), pages 913–940.

[14] Lionel C Briand, Yvan Labiche, HD Yan, and Massimiliano Di Penta. "A controlled experiment on the impact of the object constraint language in UML-based maintenance". In: *International Conference on Software Maintenance*. IEEE. IEEE, 2004, pages 380–389.

[15] Mara Victoria Cengarle and Alexander Knapp. "Towards OCL/RT". In: *International Symposium of Formal Methods Europe on Formal Methods - Getting IT Right*. Springer-Verlag, 2002, pages 390–409. DOI: 10.1007/3-540-45614-7_22.

[16] Andreas Classen, Patrick Heymans, Pierre-Yves Schobbens, Axel Legay, and Jean-François Raskin. "Model checking lots of systems: efficient verification of temporal properties in software product lines". In: *International Conference on Software Engineering (ICSE 2010)*. 2010, pages 335–344. DOI: 10.1145/1806799.1806850.

[17] Alcino Cunha. "Bounded Model Checking of Temporal Formulas with Alloy". In: *Abstract State Machines, Alloy, B, TLA, VDM, and Z*. Volume 8477. LNCS. Springer, 2014, pages 303–308.

[18] Krzysztof Czarnecki and Michal Antkiewicz. "Mapping Features to Models: A Template Approach Based on Superimposed Variants". In: *International Conference on Generative Programming and Component Engineering GPCE 2005*. 2005, pages 422–437.

[19] Krzysztof Czarnecki, Paul Grünbacher, Rick Rabiser, Klaus Schmid, and Andrzej Wąsowski. "Cool features and tough decisions: a comparison of variability modeling approaches". In: *Variability Modelling for Model-Driven Development of Software Product Lines (VaMoS)*. ACM, 2012, pages 173–182.

[20] Krzysztof Czarnecki, Simon Helsen, and Ulrich W. Eisenecker. "Formalizing cardinality-based feature models and their specialization". In: *Software Process: Improvement and Practice* 10.1 (2005), pages 7–29.

[21] Krzysztof Czarnecki and Andrzej Wąsowski. "Feature Diagrams and Logics: There and Back Again". In: *International Software Product Line Conference, SPLC*. IEEE Computer Society, 2007, pages 23–34.

[22] Werner Damm and David Harel. "LSCs: Breathing life into message sequence charts". In: *Formal methods in system design* 19.1 (2001), pages 45–80.

[23] Alexandre David, Kim Guldstrand Larsen, Axel Legay, Ulrik Nyman, and Andrzej Wąsowski. "ECDAR: An Environment for Compositional Design and Analysis of Real Time Systems". In: *ATVA*. Volume 6252. LNCS. Springer, 2010, pages 365–370.

[24] P Dissaux, Jean-Paul Bodeveix, M Filali, P Gaufillet, and F Vernadat. "AADL behavioral annex". In: *Proceedings of DASIA conference, Berlin*. 2006.

[25] Matthew B. Dwyer, George S. Avrunin, and James C. Corbett. "Patterns in Property Specifications for Finite-state Verification". In: *International Conference on Software Engineering (ICSE 1999)*. 1999, pages 411–420.

[26] Uli Fahrenberg, Mathieu Acher, Axel Legay, and Andrzej Wasowski. "Sound Merging and Differencing for Class Diagrams". In: *Fundamental Approaches to Software Engineering - 17th International Conference, FASE 2014, Held as Part of the European Joint Conferences on Theory and Practice of Software, ETAPS 2014, Grenoble, France, April 5-13, 2014, Proceedings*. Edited by Stefania Gnesi and Arend Rensink. Volume 8411. Lecture Notes in Computer Science. Springer, 2014, pages 63–78. ISBN: 978-3-642-54803-1. DOI: 10.1007/978-3-642-54804-8_5.

[27] Peter H Feiler, David P Gluch, and John J Hudak. *The architecture analysis & design language (AADL): An introduction*. Technical report CMU/SEI-2006-TN-011. Pittsburgh, Pennsylvania: Software Engineering Institute, Carnegie Mellon University., 2006.

[28] Dario Fischbein, Sebastian Uchitel, and Victor A. Braberman. "A foundation for behavioural conformance in software product line architectures". In: *Workshop on Role of software architecture for testing and analysis ROSATEA*. ACM, 2006, pages 39–48.

[29] Stephan Flake and Wolfgang Mueller. "Formal Semantics of Static and Temporal State-Oriented OCL Constraints". In: *Software & Systems Modeling (SOSYM)* 2.3 (Oct. 2003), pages 164–186.

[30] Marcelo F. Frias, Juan P. Galeotti, Carlos López Pombo, and Nazareno Aguirre. "DynAlloy: upgrading Alloy with actions". In: *International Conference on Software Engineering (ICSE 2005)*. 2005, pages 442–451.

[31] Sanford Friedenthal, Alan Moore, and Rick Steiner. *A practical guide to SysML: the systems modeling language*. Morgan Kaufmann, 2014.

[32] Jesús Padilla Gaeta and Krzysztof Czarnecki. "Modeling aerospace systems product lines in SysML". In: *International Software Product Line Conference, SPLC*. ACM, 2015, pages 293–302.

[33]  Javier González-Huerta, Silvia Abrahão, and Emilio Insfran. "Automatic derivation of AADL product architectures in software product line development". In: *Proc. of the 1st Architecture Centric Virtual Integration Workshop. Valencia: CEUR*. 2014, pages 69–78.

[34]  Iris Groher and Markus Völter. "Aspect-Oriented Model-Driven Software Product Line Engineering". In: *Transactions on Aspect-Oriented Software Development VI*. Lecture Notes in Computer Science 5560 (2009), pages 111–152. DOI: 10.1007/978-3-642-03764-1_4.

[35]  Arne Haber, Carsten Kolassa, Peter Manhart, Pedram Mir Seyed Nazari, Bernhard Rumpe, and Ina Schaefer. "First-class variability modeling in Matlab/Simulink". In: *Variability Modelling for Model-Driven Development of Software Product Lines (VaMoS)*. ACM, 2013, 4:1–4:8.

[36]  David Harel. "Statecharts: A Visual Formalism for Complex Systems". In: *Sci. Comput. Program.* 8.3 (1987), pages 231–274.

[37]  David Harel, Asaf Kleinbort, and Shahar Maoz. "S2A: A compiler for multi-modal UML sequence diagrams". In: *Fundamental Approaches to Software Engineering*. Springer, 2007, pages 121–124.

[38]  David Harel, Shahar Maoz, Smadar Szekely, and Daniel Barkan. "PlayGo: towards a comprehensive tool for scenario based programming". In: *ASE*. ACM. 2010, pages 359–360.

[39]  Oystein Haugen, Andrzej Wąsowski, and Krzysztof Czarnecki. "CVL: Common Variability Language". In: *International Software Product Line Conference, SPLC 2103*. ACM, 2013, page 277.

[40]  Gerard J. Holzmann. *Design and Validation of Computer Protocols*. Prentice-Hall, Inc., 1991. ISBN: 0-13-539925-4.

[41]  John Hutchinson, Jon Whittle, Mark Rouncefield, and Steinar Kristoffersen. "Empirical assessment of MDE in industry". In: *International Conference on Software Engineering (ICSE 2011)*. 2011, pages 471–480.

[42]  Daniel Jackson. *Software Abstractions: logic, language, and analysis. Revised edition*. MIT press, Nov. 2011.

[43]  Kyo Kang, Sholom Cohen, James Hess, William Novak, and A. Peterson. *Feature-Oriented Domain Analysis (FODA) Feasibility Study*. Technical Report CMU/SEI-90-TR-021. Software Engineering Institute, Carnegie Mellon University, 1990.

[44]  Bilal Kanso and Safouan Taha. "Specification of temporal properties with OCL". In: *Science of Computer Programming* 96, Part 4 (2014), pages 527–551.

[45]  Stuart Kent, Andy Evans, and Bernhard Rumpe. "UML Semantics FAQ". In: *ECOOP'99 Workshop Reader*. Springer, 1999, pages 33–56. ISBN: 978-3-540-46589-8.

[46]  John Krogstie. "Evaluating UML Using a Generic Quality Framework". In: *UML and the Unified Process*. Edited by Liliana Favre. Hershey, PA, USA: IRM Press, Apr. 1, 2003, pages 1–22. ISBN: 978-1-931777-44-5.

[47] Kim Guldstrand Larsen, Ulrik Nyman, and Andrzej Wąsowski. "Modal I/O Automata for Interface and Product Line Theories". In: *ESOP*. Volume 4421. LNCS. Springer, 2007, pages 64–79.

[48] Kim Guldstrand Larsen, Ulrik Nyman, and Andrzej Wąsowski. "Modeling software product lines using color-blind transition systems". In: *International Journal on Software Tools for Technology Transfer* 9.5-6 (Oct. 2007), pages 471–487.

[49] Mary E. S. Loomis, Ashwin V. Shah, and James E. Rumbaugh. "An Object Modelling Technique for Conceptual Design". In: *ECOOP*. 1987, pages 192–202.

[50] Ivano Malavolta, Patricia Lago, Henry Muccini, Patrizio Pelliccione, and Antony Tang. "What industry needs from architectural languages: An industrial survey". In: *IEEE Transactions on Software Engineering* 39.6 (June 2013), pages 869–891.

[51] Stefan Mann and Georg Rock. "Dealing with variability in architecture descriptions to support automotive product lines: Specification and analysis methods". In: *Proceedings Embedded World Conference'09*. 2009, pages 3–5.

[52] Assaf Marron and Smadar Szekely. *LSC Language Reference Manual*. Department of Computer Science and Applied Mathematics, Weizmann Institute of Science. 2014.

[53] Raghava Rao Mukkamala, Thomas T. Hildebrandt, and Janus Boris Tøth. "The Resultmaker Online Consultant: From Declarative Workflow Management in Practice to LTL". In: *ECOCW'08*. IEEE Computer Society, 2008, pages 135–142.

[54] Alexandr Murashkin. "Automotive Electronic/Electric Architecture Modeling, Design Exploration and Optimization using Clafer". Master's thesis. University of Waterloo, 2014. HDL: 10012/8780.

[55] Sarah Nadi and Stefan Krüger. "Variability Modeling of Cryptographic Components: Clafer Experience Report". In: *Variability Modelling for Model-Driven Development of Software Product Lines (VaMoS)*. ACM, 2016, pages 105–112.

[56] Juliana Alves Pereira, Kattiana Constantino, and Eduardo Figueiredo. "A Systematic Literature Review of Software Product Line Management Tools". In: *ICSR*. Volume 8919. LNCS. Springer, 2015, pages 73–89.

[57] Amir Pnueli. "The Temporal Logic of Programs". In: *18th Annual Symposium on Foundations of Computer Science*. IEEE. IEEE Computer Society, 1977, pages 46–57. DOI: 10.1109/SFCS.1977.32.

[58] Klaus Pohl, Günter Böckle, and Frank van der Linden. *Software Product Line Engineering - Foundations, Principles, and Techniques*. Springer, 2005.

[59] Rodrigo Queiroz, Leonardo Teixeira Passos, Marco Tulio Valente, Claus Hunsen, Sven Apel, and Krzysztof Czarnecki. "The shape of feature code: an analysis of twenty C-preprocessor-based systems". In: *Software & Systems Modeling (SOSYM)* 16.1 (2017), pages 77–96.

[60]   T. Dierks and E. Rescorla. *The Transport Layer Security (TLS) Protocol Version 1.2*. RFC 5246 (Proposed Standard). RFC. Updated by RFCs 5746, 5878, 6176, 7465, 7507, 7568, 7627, 7685, 7905, 7919. Fremont, CA, USA: RFC Editor, Aug. 2008. DOI: 10.17487/RFC5246. URL: https://www.rfc-editor.org/rfc/rfc5246.txt.

[61]   I. Fette and A. Melnikov. *The WebSocket Protocol*. RFC 6455 (Proposed Standard). RFC. Updated by RFCs 7936, 8307. Fremont, CA, USA: RFC Editor, Dec. 2011. DOI: 10.17487/RFC6455. URL: https://www.rfc-editor.org/rfc/rfc6455.txt.

[62]   Jordan Ross. "Synthesis and Exploration of Multi-Level, Multi-Perspective Architectures of Automotive Embedded Systems". Masters. University of Waterloo, Aug. 2016, page 165. HDL: 10012/10632.

[63]   Jordan Ross, Alexandr Murashkin, Jia Hui Liang, Michał Antkiewicz, and Krzysztof Czarnecki. "Synthesis and Exploration of Multi-Level, Multi-Perspective Architectures of Automotive Embedded Systems". In: *Software & Systems Modeling (SOSYM)* (2017).

[64]   Ina Schaefer. "Variability Modelling for Model-Driven Development of Software Product Lines". In: *VAMOS'10*. 2010, pages 85–92.

[65]   Simon Schauss, Ralf Lämmel, Johannes Härtel, Marcel Heinz, Kevin Klein, Lukas Härtel, and Thorsten Berger. "A chrestomathy of DSL implementations". In: *Proceedings of the 10th ACM SIGPLAN International Conference on Software Language Engineering, SLE 2017, Vancouver, BC, Canada, October 23-24, 2017*. Edited by Benoit Combemale, Marjan Mernik, and Bernhard Rumpe. ACM, 2017, pages 103–114.

[66]   Pourya Shaker, Joanne M. Atlee, and Shige Wang. "A feature-oriented requirements modelling language". In: *RE'12*. 2012, pages 151–160.

[67]   Shin'ichi Shiraishi. "An AADL-based approach to variability modeling of automotive control systems". In: *MODELS'10*. Springer, 2010, pages 346–360.

[68]   Amirhossein Vakili and Nancy A. Day. "Temporal Logic Model Checking in Alloy". In: *ABZ'12*. 2012, pages 150–163.

[69]   Jos B. Warmer and Anneke G. Kleppe. *The Object Constraint Language: Precise Modeling with UML*. Addison-Wesley, 1999.

[70]   Andrzej Wąsowski. "Automatic Generation of Program Families by Model Restrictions". In: *International Software Product Line Conference, SPLC 2004*. 2004, pages 73–89.

[71]   Andrzej Wąsowski. "On efficient program synthesis from statecharts". In: *Proceedings of the 2003 Conference on Languages, Compilers, and Tools for Embedded Systems (LCTES'03)*. ACM, 2003, pages 163–170.

[72]   Paul Ziemann and Martin Gogolla. "An Extension of OCL with Temporal Logic". In: *Critical Systems Development with UML*. 2002, pages 53–62.

## About the authors

**Paulius Juodisius** is a senior software engineer at Oyster.com, a TripAdvisor company. He finished his MSc in Software Development from IT University of Copenhagen in 2014 where his research focus was model driven development and various projects related to Clafer modeling language. Before joining TripAdvisor in 2015, he was a full-stack engineer for a Copenhagen based company Eff Consulting and research developer at IT University of Copenhagen. He received BSc in Software Engineering from Vilnius University in 2008.

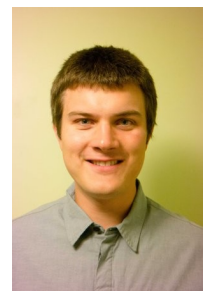

**Atrisha Sarkar** is a Ph.D. student in the David R Cheriton School of Computer Science at the University of Waterloo. She works in the Waterloo Intelligent Systems Engineering (WISE) laboratory under the supervision of Prof. Krzysztof Czarnecki. She received her masters in Computer Science from University of Waterloo in 2016, where she worked on applying machine learning techniques to software performance prediction. Before coming to Waterloo, she spent eight years in industry developing enterprise software, mostly at the IBM India Software Labs. She completed her bachelor of technology in Computer Science from Vellore Institute of Technology, India. Her ongoing Ph.D. research is on urban autonomous vehicles, artificial intelligence, and using reinforcement learning for behavioral decision making. She is a part of the *autonomoose* self-driving car project at the University of Waterloo.

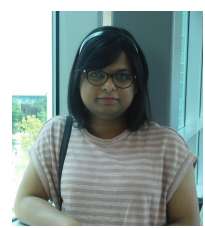

**Raghava Rao Mukkamala** currently works in computational social science using an interdisciplinary approach, combining formal modelling approaches with data-mining/machine-learning techniques, for modelling of social science phenomena in digital transformation of organisations and society. He also works with blockchain-based technologies for social business and IoT. His research work is further supported by 10 + years of professional experience in the Danish IT industry as a senior software engineer and consultant.

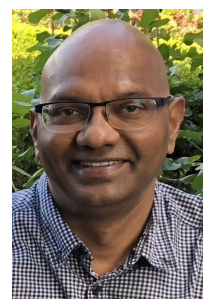

Raghava is an associate professor at department of Digitalization in Copenhagen Business School, Denmark and also an associate professor at department of Technology, Kristiania University College, Oslo, Norway. He holds a PhD degree and a MSc degree from IT University of Copenhagen, Denmark. He can be reached at rrm.digi@cbs.dk.

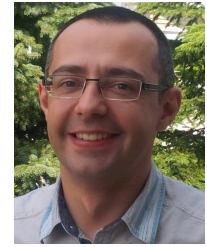

**Michał Antkiewicz** is a research engineer at the Electrical and Computer Engineering Department at the University of Waterloo. During Ph.D., Michał worked on topics related to modeling in product-line engineering (cardinality-based feature modeling and feature-based model templates) and round-trip engineering between the code using a framework API and models describing API usage (framework-specific modeling languages). Since 2011, Michał is a maintainer of Clafer Tools, including the Clafer compiler. Since 2016, Michał works on Waterloo Autonomous car project http://autonomoose.net.

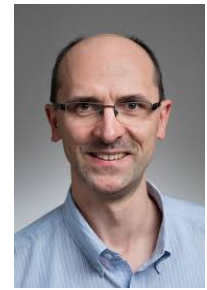

**Krzysztof Czarnecki** is a Professor of Electrical and Computer Engineering at the University of Waterloo. Before coming to Waterloo, he was a researcher at DaimlerChrysler Research (1995-2002), Germany, focusing on improving software development practices and technologies in enterprise, automotive, and aerospace sectors. While at Waterloo, he held the NSERC/Bank of Nova Scotia Industrial Research Chair in Requirements Engineering of Service-oriented Software Systems (2008-2013) and has worked on methods and tools for engineering complex software-intensive systems. He received the Premier's Research Excellence Award in 2004 and the British Computing Society in Upper Canada Award for Outstanding Contributions to IT Industry in 2008. He has also received seven Best Paper Awards, two ACM Distinguished Paper Awards, and one Most Influential Paper Award. His current research focuses on autonomous driving and the safety of systems that rely on artificial intelligence. As part of this research, he co-leads the development of UW Moose, Canada's first self-driving research vehicle (http://autonomoose.net).

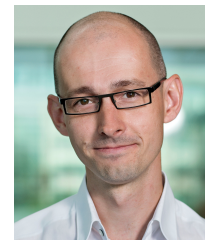

**Andrzej Wąsowski** works with design and use of technologies that improve quality of software, including issues such as correctness and maintainability. He has worked extensively with software product line methods—ways to develop software for similar products at lower cost but with higher quality. He has collaborated with open source projects (Linux kernel and ROS among others) and with industry (for example with Danfoss). Currently, he is investigating quality assurance methods for robotics platforms, in the H2020 project ROSIN.

Andrzej Wasowski is a professor of Software Engineering at IT University in Copenhagen (ITU). He holds an MSc degree from Warsaw University of Technology and a PhD degree from ITU. He has previously held visiting positions at Aalborg University (Denmark), INRIA Rennes (France) and University of Waterloo (Canada).